\documentclass{article}

\usepackage{arxiv}

\usepackage[utf8]{inputenc} 
\usepackage[T1]{fontenc}    
\usepackage{hyperref}       
\usepackage{url}            
\usepackage{booktabs}       
\usepackage{amsfonts}       
\usepackage{nicefrac}       
\usepackage{microtype}      
\usepackage{lipsum}
\usepackage{graphicx}
\usepackage{algorithm}
\usepackage{algorithmic}

\title{Extending Isolation Forest for Anomaly Detection in Big Data via K-Means}

\author{
  Md Tahmid Rahman Laskar\\
Information Retrieval \& Knowledge Management Research Lab, York University, Canada\\
 \texttt{tahmedge@cse.yorku.ca} \\

 \And Jimmy Huang \\
Information Retrieval \& Knowledge Management Research Lab, York University, Canada \\
  \texttt{jhuang@yorku.ca} \\

 \And Vladan Smetana \\
iSecurity Consulting Inc, Canada \\
  \texttt{vladan.smetana@isecurityconsulting.com} \\
  
 \And Chris Stewart \\
iSecurity Consulting Inc, Canada \\
    \texttt{chris.stewart@isecurityconsulting.com} \\
  
 \And Kees Pouw \\
iSecurity Consulting Inc, Canada \\
      \texttt{kees.pouw@isecurityconsulting.com} \\
  
 \And Aijun An \\
York University, Canada \\
    \texttt{aan@cse.yorku.ca} \\
 \And Stephen Chan \\
Dapasoft Inc, Canada \\
\texttt{schan@dapasoft.com} \\
  
 \And Lei Liu \\
Information Retrieval \& Knowledge Management Research Lab, York University, Canada \\
  \texttt{lliu@cse.yorku.ca} \\
  
}

\begin{document}
\maketitle

\begin{abstract}
Industrial Information Technology (IT) infrastructures are often vulnerable to cyberattacks. To ensure security to the computer systems in an industrial environment, it is required to build effective intrusion detection systems to monitor the cyber-physical systems (e.g., computer networks) in the industry for malicious activities. This paper aims to build such intrusion detection systems to protect the computer networks from cyberattacks. More specifically, we propose a novel unsupervised machine learning approach that combines the K-Means algorithm with the Isolation Forest for anomaly detection in industrial big data scenarios. Since our objective is to build the intrusion detection system for the big data scenario in the industrial domain, we utilize the Apache Spark framework to implement our proposed model which was trained in large network traffic data (about 123 million instances of network traffic) stored in Elasticsearch. Moreover, we evaluate our proposed model on the live streaming data and find that our proposed system can be used for real-time anomaly detection in the industrial setup. In addition, we address different challenges that we face while training our model on large datasets and explicitly describe how these issues were resolved. Based on our empirical evaluation in different use-cases for anomaly detection in real-world network traffic data, we observe that our proposed system is effective to detect anomalies in big data scenarios. Finally, we evaluate our proposed model on several academic datasets to compare with other models and find that it provides comparable performance with other state-of-the-art approaches. 
\end{abstract}

\keywords{Big Data \and Cyber Security \and Network Intrusion Detection \and Machine Learning \and Apach Spark}

\section{Introduction}
Anomalies in network traffic are the patterns in data that deviate significantly from the expected ones \cite{chandola2009anomalysurvey,ahmed2016anomalysurvey,chalapathy2019deepsurvey,wu2020anomaly}. Since the computer networks in a cyber-physical system are often vulnerable to cyberattacks which can severely disrupt the overall system, it is required to identify the malicious events in the network traffic data in such systems \cite{leung2005unsupervisedclustering,terzi2017bigdata}. Thus, a reliable network intrusion detection system is needed that should be robust enough so that the novel anomalies that did not appear before could be identified effectively. In addition, these systems should also detect the anomalies efficiently so that they can be applied to the live streaming data. In order to develop such systems, various approaches are utilized in recent years which include but not limited to the following: 
\begin{itemize}
    \item Rule-Based: where a set of rules are manually given and the system detects anomalies based on these rules.
    \item Supervised Learning: where a set of labeled training data is given and various machine learning algorithms are used to learn a prediction model from these labeled data.
    \item Unsupervised Learning: where the machine learning models learn to identify anomalies from the unlabeled data.
    \item Reinforcement Learning: where machine learning algorithms are trained to make a sequence of decisions based on reward and punishment. 
\end{itemize}
 
However, there are several challenges when building anomaly or intrusion detection systems using the approaches mentioned above \cite{chandola2009anomalysurvey}. For example, the rule-based approaches fail to recognise the malicious events for which no rules have been specified \cite{thottan2003anomalyIP}. Thus, such rule-based systems are restricted to only recognize events for which rules are already given. Similar to the rule-based approach, the supervised approach also has several limitations where one of the major limitations is that it requires labeled training data \cite{chandola2009anomalysurvey}. Thus, supervised learning algorithms are not applicable in scenarios when labeled training data are not available \cite{chandola2009anomalysurvey,ahmed2016anomalysurvey,choi2019unsupervisedautoencoder}. Since there is a lack of labeled datasets to train network anomaly detection models, it is required to build such systems with limited supervision or unlabeled training data \cite{chandola2009anomalysurvey,ahmed2016anomalysurvey,choi2019unsupervisedautoencoder,otoum2018adaptively}. As the unlabeled training datasets are easier to obtain since such datasets do not require any human intervention for data labeling, various unsupervised algorithms have been used in recent years in scenarios when the labeled training data are not present \cite{choi2019unsupervisedautoencoder,amer2013enhancingunsupervisedSVM,ester1996dbscan,liu2008isolation,liu2012isolation}. Nonetheless, training an unsupervised method to learn patterns in data is more challenging compared to its supervised counterpart \cite{ahmed2016anomalysurvey}. Moreover, most of the unsupervised machine learning models proposed for anomaly detection are also evaluated only on relatively smaller datasets \cite{choi2019unsupervisedautoencoder,liu2008isolation,ester1996dbscan} and their evaluation in the real world large-scale industrial settings when labeled training data are not available is yet to be investigated. Additionally, some of the prior work utilized reinforcement learning for intrusion detection \cite{otoum2019empowering,tiyas2014reinforced,xu2005reinforcement,servin2005multireinforcement}. However, such models based on reinforcement learning are also not evaluated in big data scenarios yet. To this end, we aim to address the above challenges to build an effective intrusion detection system for big data scenarios. Our motivations in this paper are discussed in the following section.

\subsection{Motivation} It is to be noted that the Apache Spark\footnote{ \url{https://spark.apache.org/}} framework which is utilized for large-scale data processing includes several machine learning algorithms through its MLlib\footnote{\url{https://spark.apache.org/mllib/}} library \cite{meng2016mllib}. Thus, in circumstances when labeled training data are not available, the unsupervised learning algorithms in Apache Spark can be utilized. Nonetheless, various state-of-the-art machine learning models, such as Isolation Forest \cite{liu2008isolation}, Auto Encoders \cite{choi2019unsupervisedautoencoder}, and DBSCAN \cite{ester1996dbscan}, as well as reinforcement learning algorithms such as QL-IDS \cite{otoum2020comparative} are not included in the MLlib library yet. Thus, most prior work that addressed the anomaly detection problem in big data scenarios did not utilize the state-of-the-art anomaly detection algorithms due to the absence of such models in the Apache Spark framework \cite{ahmed2016anomalysurvey,chalapathy2019deepsurvey}. To address these challenges, in this paper, we propose a novel unsupervised intrusion detection system that can be effectively utilized to detect the anomalies or intrusions in large industrial network traffic data when labeled training data are not available. Due to the effectiveness of Isolation Forest compared to the other unsupervised approaches \cite{liu2008isolation,domingues2018comparative}, we are motivated to utilize it to implement an intrusion detection system that can detect intrusions on large network traffic data. For that purpose, we propose a novel approach via combining Isolation Forest with K-Means that extends the original Isolation Forest algorithm for anomaly detection in big data. Contrary to the prior work that evaluated Isolation Forest only on comparatively smaller-sized academic datasets \cite{liu2008isolation,hariri2018extendedIF}, we primarily evaluate our proposed model on a large industrial dataset to demonstrate its effectiveness in big data scenarios. In the following section, our contributions in this paper are demonstrated.   


\subsection{Contributions}

Our major contributions presented in this paper are as follows:

\begin{itemize}
    \item First, in contrary to the previously proposed anomaly detection systems which were only evaluated on smaller-sized academic datasets \cite{liu2008isolation,ester1996dbscan}, our objective in this paper is to propose an intrusion detection system that can learn from the huge amount of unlabeled data in industries so that the trained model can detect intrusions in real-world large network traffic data. For this purpose, we leverage the Apache Spark \cite{titicacasparkiforest} framework to implement our proposed system that we train in an industrial network traffic dataset containing more than 123 million instances where the size of the dataset is above 1 terabyte (TB).

    \item Second, we address the following fundamental issues in Isolation Forest: (i) it requires information about the ratio of anomalies (i.e., contamination ratio) in the training data while providing this information may require the construction of a manually labeled training dataset which is a very time-consuming process in big data scenarios \cite{liu2008isolation,titicacasparkiforest,linkedin}; and (ii) it needs to find out a threshold to convert the anomaly scores predicted by the model to different labels \cite{liu2008isolation,titicacasparkiforest,linkedin}, which is a tedious process that becomes more expensive in large datasets. For that purpose, we propose a novel approach that combines the K-Means algorithm with Isolation Forest for anomaly detection in big data scenarios. We show that our proposed model that utilizes fewer parameters than the Isolation Forest algorithm is not only applicable in real-world big data scenarios in industries, but also more effective for anomaly detection than the original Isolation Forest model. Additionally, we further analyze our findings through extensive experiments in 12 academic datasets.
    
    \item Finally, we extend our trained model to detect intrusions on the live steaming data in the industrial setting and show that our proposed system can also detect intrusions in real-time both effectively and efficiently. To the best of our knowledge, this is the first work in which a model based on isolation forest has been evaluated on the industrial streaming data for real-time detection of anomalies \cite{habeeb2019realstreamingsurvey}. As a secondary contribution: to reproduce our experiments in big data scenarios, we extensively describe how we mitigate various limitations in the existing Spark-based implementation of Isolation Forest \cite{titicacasparkiforest} that prevent this model from being trained on large datasets.
\end{itemize}


\section{Related Work}

In the earlier years, various rule-based techniques were utilized to develop anomaly detection systems \cite{ahmed2016anomalysurvey,chandola2009anomalysurvey}. 
Although the rule-based systems are effective to detect the known anomalies for which rules are given, they are slow to detect anomalies in real-time and also fail to adapt to the evolving network environment \cite{thottan2003anomalyIP}. Moreover, the detection rate of anomalies using different rule-based approaches relies heavily on network administrators' expertise and domain knowledge about the known attacks \cite{thottan2003anomalyIP}. Thus, unknown attacks for which no rules are given by the network experts may remain undetected by such systems. To alleviate these limitations of rule-based systems, various machine learning-based algorithms have been proposed in recent years for network intrusion detection \cite{ahmed2016anomalysurvey,chandola2009anomalysurvey,liu2008isolation,feng2014mining}. In such machine learning-based systems, usually two approaches were mostly used: i) supervised learning approach and ii) unsupervised learning approach \cite{mukherjee2012intrusion,liao2002use,ester1996dbscan,choi2019unsupervisedautoencoder,aloqaily2019intrusiondecision,liu2008isolation,sommer2010outsideML,otoum2020comparative,otoum2018adaptively}. In addition to these two approaches, some of the prior work also utilized reinforcement learning for anomaly detection \cite{otoum2019empowering,tiyas2014reinforced,xu2005reinforcement,servin2005multireinforcement}. Among various machine learning-based approaches, most previous work relied on supervised learning that required labeled training data of normal and abnormal instances to train the models \cite{ahmed2016anomalysurvey,chandola2009anomalysurvey,otoum2020comparative}. 
For the supervised training, various classification-based machine learning algorithms such as Support Vector Machine  \cite{mukkamala2002intrusion,sotiris2010anomalySVMbayesian,JH7}, Naive Bayes \cite{mukherjee2012intrusion}, Random Forest \cite{farnaaz2016random,otoum2017detectionrandom}, K-Nearest Neighbour \cite{liao2002use}, Decision Tree \cite{aloqaily2019intrusiondecision}, as well as ensemble learning methods that combine multiple classifiers together \cite{otoum2020novelensemble} are used for anomaly or intrusion detection. Some work also combined different learning algorithms together \cite{corsini2006combining,feng2014mining}. For example, Feng et al. \cite{feng2014mining} combined the supervised Support Vector Machine model with the unsupervised Self-Organized Ant Colony Network (CSOACN) \cite{feng2006csoacn} architecture and proposed the CSVAC model for network intrusion detection. Similar to the most prior work on anomaly detection mentioned above, they also used labeled training data to train their CSVAC model.

Despite the effectiveness of supervised learning models, they cannot be utilized in many scenarios. For instance, constructing labeled datasets, especially in big data scenarios is a costly process. Since the network experts need to manually label the normal and abnormal instances in network traffic to create the labeled dataset \cite{choi2019unsupervisedautoencoder}, the task of manual labeling becomes very expensive in big data scenarios. In such circumstances when labeled training data are not available, the supervised learning algorithms cannot be utilized. To tackle such scenarios where only unlabeled training data are available, the unsupervised learning algorithms are usually applied \cite{choi2019unsupervisedautoencoder,tian2014anomalysupervisedlabel}. For unsupervised learning, various clustering-based machine learning models \cite{leung2005unsupervisedclustering,feng2006csoacn,ester1996dbscan} have been utilized for anomaly detection which can be divided into several categories, such as: partitioning method, distance-based method, or density-based method. However, these models have several disadvantages. For instance, the partitioning methods are sensitive to the width of the cluster that requires the repetition of experiments to choose the optimal width \cite{leung2005unsupervisedclustering}. Note that this repetition of experiments becomes a very time-consuming process
in large datasets.
Moreover, for other clustering models which are based on distance or density measures \cite{leung2005unsupervisedclustering,ester1996dbscan}, the computational cost of measuring the distance is expensive \cite{liu2008isolation}. In recent years, the successful utilization of deep learning techniques to solve various computing problems has also inspired researchers to apply deep learning for anomaly detection \cite{choi2019unsupervisedautoencoder,ergen2019unsupervisedlstm,mukkamala2002intrusion,vinayakumar2017cnn,shone2018deep1,javaid2016deep2,yin2017deep3,otoum2020comparative}. However, similar to the various machine learning algorithms, these deep learning-based models were also mostly evaluated on relatively smaller datasets and their evaluation on large datasets in the industrial domain is yet to be investigated.

To overcome the above limitations of several anomaly detection models, Liu et al. \cite{liu2008isolation} proposed a tree structured anomaly detection approach: the Isolation Forest. In comparison to the clustering-based machine learning models as well as the deep learning-based models used for anomaly detection, the Isolation Forest algorithm has several advantages over both of them. While comparing with the clustering-based machine learning models \cite{leung2005unsupervisedclustering}, the Isolation Forest does not require the calculation of distance or density measures in high-dimensional data to detect the anomalies. Thus, it eliminates the major computational cost of distance calculation required in different distance-based or density-based clustering methods. Furthermore, the Isolation Forest also has a linear time complexity with low memory requirement \cite{liu2008isolation}. More importantly, it was shown by Liu et al. \cite{liu2008isolation} that the isolation forest algorithm had the capacity to scale up to handle high-dimensional problems in large datasets. This makes Isolation Forest a good choice to be utilized for anomaly detection in big data scenarios.

Though Isolation Forest has several advantages over other unsupervised approaches, it was also mostly evaluated on academic datasets only \cite{liu2008isolation,titicacasparkiforest}. To be noted that, these datasets are relatively smaller compared to the big data settings in different industries. To utilize machine learning for big data processing, Apache Spark has been widely used for such large-scale data processing \cite{zaharia2016apachespark}. It contains the MLlib library which includes several unsupervised machine learning algorithms \cite{zaharia2016apachespark}. However, the Spark MLlib does not include the Isolation Forest algorithm even though Isolation Forest is one of the most top performing approaches for anomaly detection \cite{domingues2018comparative}. Though there are some Spark-based third-party implementations of the Isolation Forest, these third-party libraries are not investigated yet to detect anomalies on a large volume of data in an industrial setting. In this work, we extend Isolation Forest via combining it with K-Means to propose a novel approach for anomaly detection in big data scenarios. Note that, our proposed approach also addresses the following fundamental issues of Isolation Forest: (i) tedious process of converting anomaly scores to different labels, and (ii) dependence on the contamination ratio in the training data that may require manual labeling of datasets which is very expensive in big data scenarios. Moreover, we also evaluate our proposed system for the real-time detection of anomalies on the live streaming data in an industrial environment.




\begin{figure*}[t!]
\begin{center}
\includegraphics[width=\linewidth]{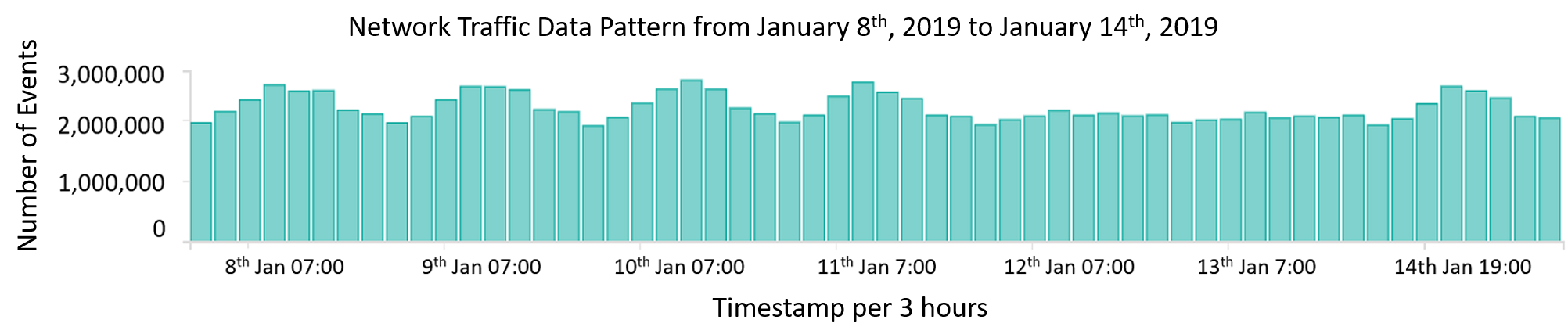}

\caption{
Normal network traffic pattern at iSecurity. Y-axis represents the total number of events occurred on each timestamp represented by the X-axis.  
}
\label{fig:Normal}
\end{center}
\end{figure*}

\section{Task Description}

In this work, we focus on building a machine learning-based intrusion detection system to provide cyber security at iSecurity\footnote{\url{https://www.isecurityconsulting.com/}}. Note that iSecurity is an IT organization that provides various security services to its clients. On January 16, 2019, a client of iSecurity was targeted by cyberattacks. Since cyberattacks are very threatening for the security and privacy of organizational data, a strong security service is required to protect the computer or network servers (i.e., cyber-physical systems) in an organization from such cyberattacks.

\begin{figure*}[t!]
\begin{center}
\includegraphics[width=\linewidth]{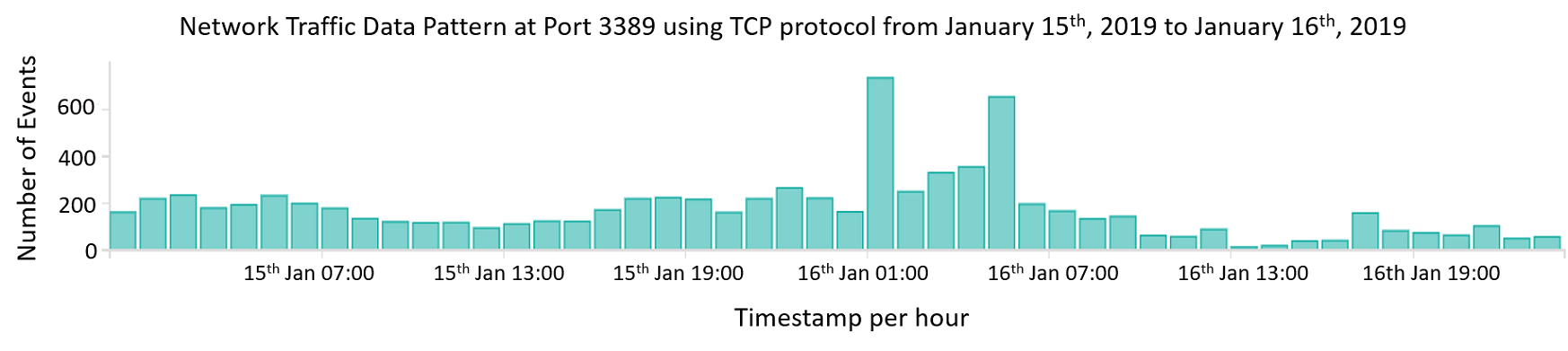}

\caption{
Network traffic pattern during a cyberattack targeting the port 3389 using the TCP protocol on January 16th: the sudden spike in the number of events at 1 to 2 am and 5 to 6 am are flagged as cyberattacks by the network expert at iSecurity. At any other time interval, there are no such spikes.
}
\label{fig:3389}
\end{center}
\end{figure*}

\begin{figure*}[t!]
\begin{center}
\includegraphics[width=\linewidth]{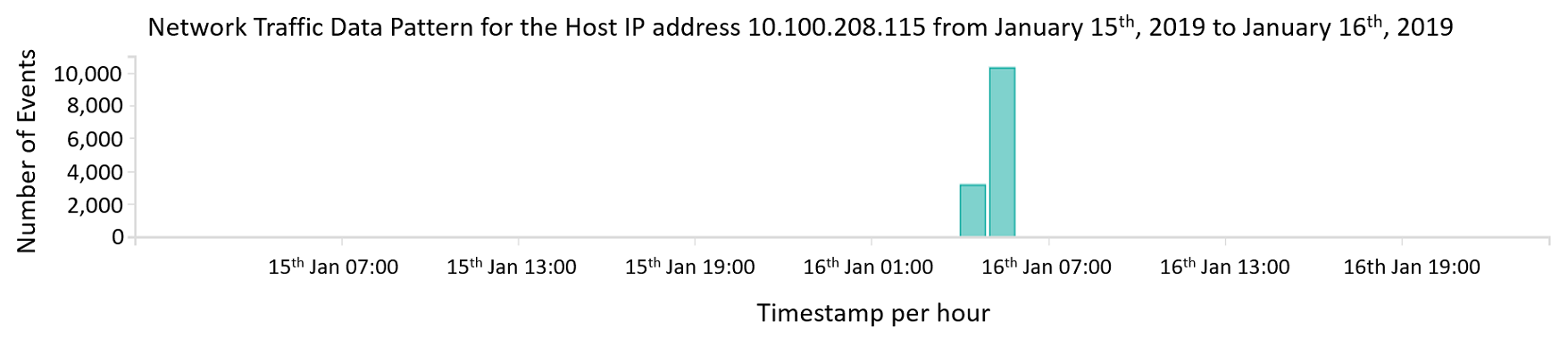}

\caption{
Network traffic pattern during a cyberattack from a specific IP address 10.100.208.115 on January 16th: the sudden spike in the number of events between 4 am and 6 am (inclusive) are flagged as cyberattacks by the network expert at iSecurity. At any other time interval, there are no such events.
}
\label{fig:IP}
\end{center}
\end{figure*}

At iSecurity, network traffic logs from client servers are collected using Graylog\footnote{\url{https://www.graylog.org/}\label{graylog}}\textsuperscript{,}\footnote{\url{https://github.com/Graylog2/graylog2-server}} and then indexed into Elasticsearch \cite{gormley2015Elasticsearch}. For the cyberattacks that occurred on January 16th, the network experts at iSecurity manually identified some incidents as potential cyberattacks. Nonetheless, manual labeling of anomaly instances in large datasets is a very tedious process. In Figure \ref{fig:Normal}, we show the one week network traffic logs at the iSecurity client where the cyberattacks occurred. It contains all the instances from January 8th to January 14th which were considered as normal traffic patterns by the network administrators at iSecurity. In Figure \ref{fig:3389} and Figure \ref{fig:IP}, we show some use-cases of cyberattacks that were identified by the network experts. Here, in Figure \ref{fig:3389}, the network traffic pattern during a cyberattack is demonstrated which originated at around 1 am on January 16th and continued till about 6 am. The attacks were occurred using the TCP protocol and targeted a specific port address 3389. Comparing with the activities on that port using the TCP protocol on any other day, we can see that there is a sudden spike in the number of connections during that time. In Figure \ref{fig:IP}, the network traffic pattern during a cyberattack from a specific IP address is shown. We can see that on January 16th (within 4 to 6 am), from the IP address 10.100.208.115, near 15000 incidents occurred. However, on any other time interval, there are no incidents happened from that IP address. 

In this paper, we aim to identify such anomalies or intrusions in large network traffic data using machine learning models due to the superior performance of machine learning-based algorithms in big data scenarios as well as when labeled datasets are not available \cite{liu2008isolation,choi2019unsupervisedautoencoder,ahmed2016anomalysurvey,chandola2009anomalysurvey,domingues2018comparative}. In the following, we first describe our proposed machine learning model: the IForest-KMeans which extends Isolation Forest for anomaly detection in big data scenarios via combining with K-Means. Then, we describe our proposed system architecture that utilizes the proposed IForest-KMeans model to detect anomalies in large network traffic data stored in Elasticsearch. 

\section{Proposed Machine Learning Model}

In our proposed model, we combine the Isolation Forest algorithm with K-Means. In this section, at first we briefly introduce readers to the K-Means and Isolation Forest algorithms. Then we discuss how we utilize these two algorithms together for anomaly detection in large datasets. 

\subsection{The K-Means Algorithm}

The K-Means algorithm tries to partition $n$ data points into $K$ clusters in such a way that each data point belongs to at most one group. The value of $k$ in K-Means is pre-defined. The algorithm tries to separate the data points in $k$ different clusters where the objective is to keep the data points that are more similar to be in the same cluster along with ensuring the distance between the data points in different clusters as distant as possible. The working mechanism of K-Means is as follows: 

\begin{enumerate}
    \item First, $k$ data points are arbitrarily chosen. Then each data point is put to one of the $k$ clusters in such a way that each cluster contains at most one point. 
    \item Then, for each cluster, the arithmetic mean of the data points in that cluster is calculated which is called as the centroid for that cluster. 
    \item Afterwards, the squared distance between all data points and the cluster centroids is calculated. Based on the value of the calculated distance, each data point is assigned to the closest cluster. 
    \item Steps 2-4 are repeated until there is no change to the centroids, i.e., the sum of the squared distance between the data points and the cluster’s centroid has reached the minimum.
    
\end{enumerate}



\subsection{The Isolation Forest Algorithm}

\begin{figure}[t!]
\begin{center}
\includegraphics[height=6cm,width=12cm]{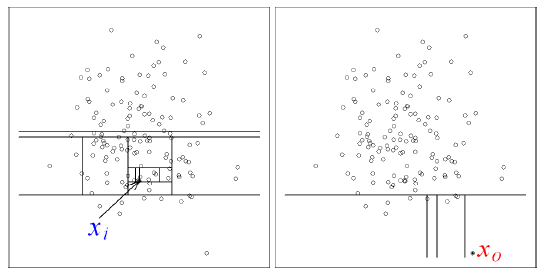}

\caption{
Splitting the data points using the Isolation Forest algorithm: (a) Isolating the normal point x\textsubscript{i} requires more splits, (b) Isolating the anomalous point x\textsubscript{0} requires less splits. This figure is taken from Liu et al. \cite{liu2008isolation}. 
}
\label{fig:IF}
\end{center}
\end{figure}


The Isolation Forest is an unsupervised anomaly detection algorithm, which comes with the intuition that the anomalies 
are few and have unusual feature values compared to the usual ones \cite{liu2008isolation,liu2012isolation}. It is a tree structured algorithm that can isolate every single instance in a dataset (see Figure \ref{fig:IF}). Since the anomalies are few and different, they are isolated closer to the root of the tree. Whereas the normal points are isolated at the deeper end of the tree. 

The Isolation Forest algorithm builds an ensemble of trees for a given set of training data. The anomalies are those instances in the dataset which have short average path lengths on these trees. To construct the Isolation Forest, at first, the number of isolation trees\footnote{Each tree in the Isolation Forest is called as isolation tree.} in the forest is required to be selected. Then, each isolation tree is created using the following steps:

\begin{enumerate}
    \item Randomly sample $n$ instances from the training dataset.
    \item Randomly choose a feature to split upon.
    \item Randomly choose a split value from a uniform distribution spanning from the minimum value to the maximum value of the feature chosen in Step 2.
\end{enumerate}

Steps 2 and 3 are repeated recursively until all $n$ instances from the randomly selected sample are isolated in leaf nodes of the isolation tree. In this way, the anomalies will require fewer random splits to be isolated in leaf nodes which will result in a shorter expected path length from the root node to the leaf node. 

Suppose, there are $n$ instances in a dataset and $h(x)$ is the path length for the instance $x$. The value of the path length $h(x)$ is then normalized using the average path length $c(n)$.
As the isolation trees have an equivalent structure to the Binary Search Tree (BST) \cite{preiss2008datastructure}, Liu et al. \cite{liu2008isolation} estimated the value of $c(n)$ to be the same as the path length of unsuccessful search in BST. They calculated the value of $c(n)$ based on the following equation where $H(i)$ is the harmonic number that can be estimated by $ln(i)$ $+$ $0.5772156649$ (Euler’s constant) \cite{liu2008isolation}:

\begin{equation}
    c(n) = 2H(n - 1) - (2(n - 1)/n)
\end{equation}
 
For a dataset having $n$ instances, the anomaly score of each data point $x$ is calculated as follows:

\begin{equation}
    s(x, n) = 2^{-({E(h(x))/c(n))}}
\end{equation}

Here, $E(h(x))$ is the average of $h(x)$ from a collection of isolation trees. The more the anomaly score of a data point is close to 1, the higher the possibility of that data point being an anomaly. On the contrary, the anomaly score closer to 0 indicates that the data point is a more of normal observation.

\subsection{The IForest-KMeans Model}
Since our goal is to build an intrusion detection system for big data scenarios, we leverage the Apache Spark framework since it is widely used for large-scale data processing. For that purpose, we ought to utilize the Spark-based implementation of Isolation Forest by Yang et al. \cite{titicacasparkiforest}. Recall that there are some fundamental issues of Isolation Forest: (i) it needs to find out a threshold to convert the anomaly scores predicted by the model to different labels which is an expensive process, and (ii) it requires information of the contamination ratio in the training data which is troublesome to provide in big data scenarios. Moreover, we also find that the existing Spark-based implementations of Isolation Forest have some issues to convert the anomaly scores to the specific labels \cite{titicacasparkiforest,linkedin}. For instance, the Spark-based Isolation Forest by Yang et al. \cite{titicacasparkiforest} requires the calculation of approximate quantile to convert the anomaly scores to the predicted labels. However, they also mentioned that calculating the exact value for approximate quantile is expensive for large datasets \cite{titicacasparkiforest}. To tackle this issue, they suggested to increase the value of the relative error while calculating the approximate quantile. Nonetheless, increasing the value of the relative error will also lead to the deterioration of performance \cite{titicacasparkiforest}. It should be pointed out that this issue also exists in another available Spark-based implementation of Isolation Forest  while converting the anomaly scores to the predicted labels \cite{linkedin}. 

\begin{figure*}[t!]
\begin{center}
\includegraphics[width=\linewidth]{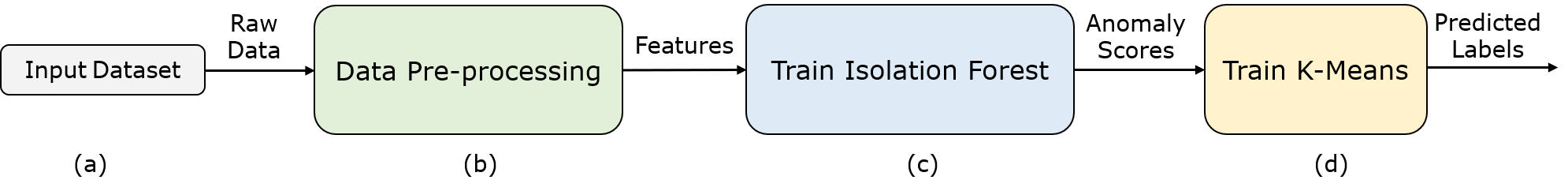}

\caption{
An overview of our proposed IForest-KMeans model. The (a) input dataset is first (b) pre-processed. Then the relevant features from that dataset are given as input to the (c) Isolation Forest model. The anomaly scores predicted by Isolation Forest is then given as input to (d) the K-Means model where K-Means partitions the anomaly scores in $K$ clusters to predict $k$ labels. 
}
\label{fig:ifkm}
\end{center}
\end{figure*}

Thus, to address these limitations, we propose a novel method via combining the K-Means algorithm with the Isolation Forest algorithm (we denote our proposed model as IForest-KMeans). Our proposed model utilized the K-Means algorithm to convert the anomaly scores predicted by the Isolation Forest model to different labels. As K-Means is an effective algorithm to partition data points into clusters, and since the MLlib library of Apache Spark already contains K-Means for large-scale data processing, we select K-Means to partition the data points into different clusters based on the anomaly scores. In our proposed model, after pre-processing the dataset, the most relevant features in the dataset are given as input to the Isolation Forest for training. After the Isolation Forest algorithm is trained, it assigns anomaly scores to each data point. Then, we give the anomaly scores of each data point as input to train the K-Means algorithm. The trained K-Means algorithm then predicts the label of each data point. In Figure \ref{fig:ifkm}, we show the overview of our proposed IForest-KMeans model.

\section{Proposed System Architecture}

\begin{figure*}[t!]
\begin{center}
\includegraphics[width=15cm,height=7.5cm]{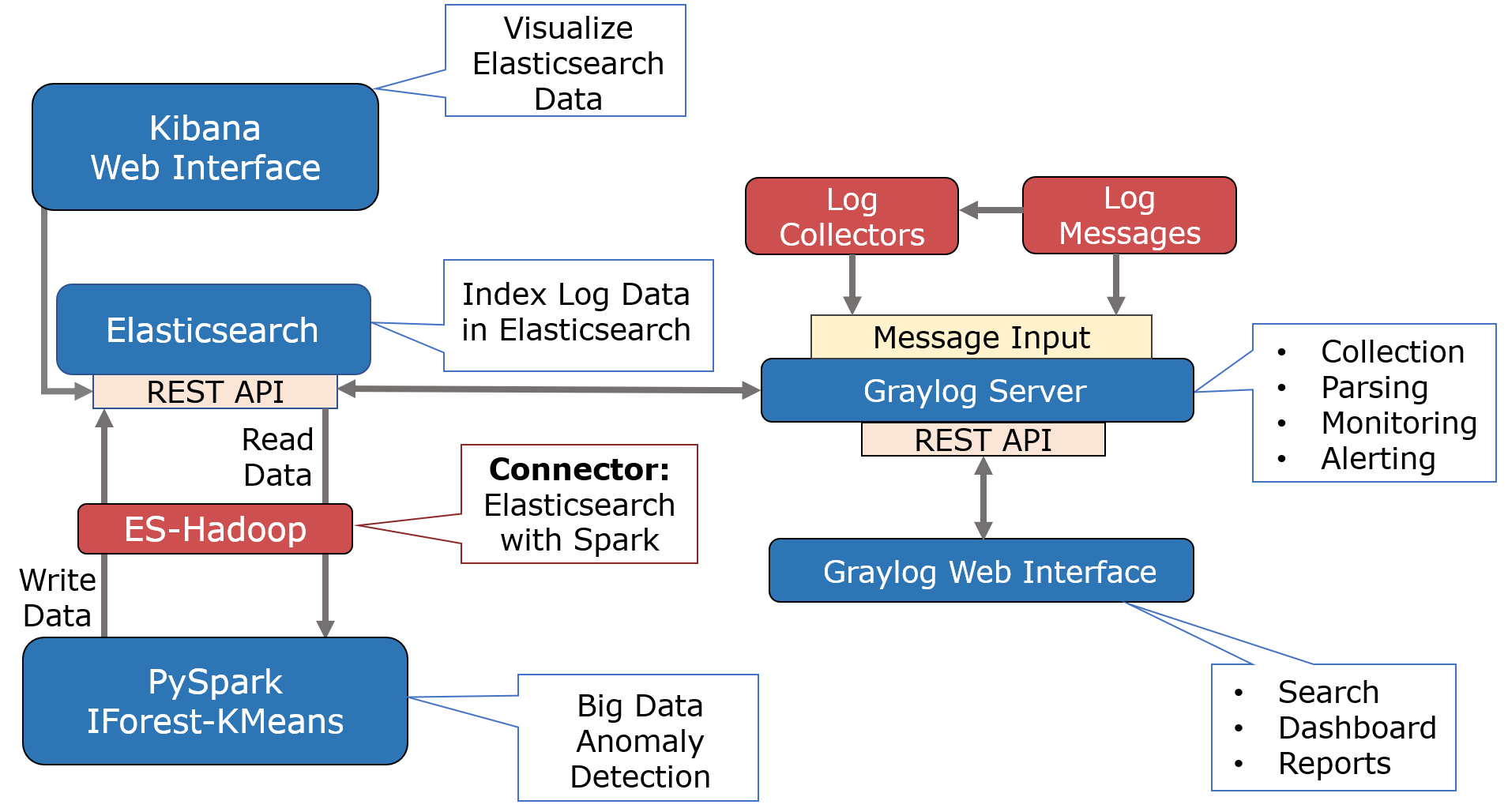}

\caption{
Proposed System Architecture for Network Intrusion Detection at iSecurity: Log data that are parsed by Graylog are stored into an Elasticsearch index. Then, the PySpark-based anomaly detection model (i.e., IForest-KMeans) utilizes the Elasticsearch-Hadoop connector to read these data, as well as write the prediction results into an Elasticsearch index to visualize using Kibana.
}
\label{fig:overallmodelarchitecture}
\end{center}
\end{figure*}

In this section, we describe our proposed system at iSecurity where we utilize our proposed IForest-KMeans model for network intrusion detection. In Figure \ref{fig:overallmodelarchitecture}, an overview of our proposed system architecture at iSecurity is shown. In the following, we describe the overall workflow of our system. 

\subsection{\textbf{Overall System Workflow}} As mentioned earlier, all the log data at iSecurity are collected and processed by Graylog. Then these log data are indexed into Elasticsearch\footnote{\url{https://www.elastic.co/}\label{elastic}}. Since the size of these log data is very large, Hadoop-based \cite{white2012hadoop} technologies are required to process them. Due to several advantages of the Apache-Spark framework over the traditional Hadoop framework \cite{zaharia2016apachespark}, we leverage the Spark technology to build our anomaly detection system. In our proposed network intrusion detection system, we read the log data from Elasticsearch using the Elasticsearch-hadoop connector to connect Apache Spark with Elasticsearch. Then the Elasticsearch-hadoop connector was utilized again to write the predicted result of our trained Spark-based anomaly detection model in the machine learning index of Elasticsearch. To analyze the data, we utilize Kibana\footnote{\url{https://www.elastic.co/kibana}\label{kibana}} \cite{gupta2015kibana} for data visualization. Below, we describe different components of our proposed system.

\subsection{\textbf{Description of System Components}} 
In this section, we describe different components in our proposed system architecture at iSecurity.

\textbf{Graylog:} Graylog provides easy management of logs. It collects event data from any data sources and then parses the data to enrich the logs. It also monitors the processing pipelines along with routing, blacklisting, modifying, and enriching messages in real-time. Furthermore, it provides the management and monitoring capabilities for the underlying Elasticsearch index.

 \textbf{Elasticsearch:} Elasticsearch is a big data solution that includes a search engine based on Lucene \cite{gormley2015Elasticsearch}. It can effectively store and index large amounts of data and provides a distributed full text search engine with RESTful API and schema free JSON document. Moreover, it provides the \textit{\textbf{Elasticsearch-hadoop}} connector that connects Spark with Elasticsearch for data processing \cite{shukla2015Elasticsearch}.  

 \textbf{Apache Spark:} Apache Spark is 
 a distributed computing engine used for processing and analyzing large amount of data \cite{zaharia2016apachespark}. Just like MapReduce in Hadoop \cite{dean2010mapreduce}, it also provides parallel processing of large datasets distributed across different clusters. However, the processing of large data on disk using MapReduce 
 is a very time-consuming process due to a large number of read/write operations.
 Spark addressed this issue by allowing the data to be processed on memory in addition to the disk, leading to a significant boost in processing speed. \cite{zaharia2016apachespark}. Moreover, for complex applications which are required to be run only on disk, Spark can process the data on disk faster than Hadoop. Since in a network system, it is required to detect anomalies in real-time which is not feasible with Hadoop, Spark provides real-time processing of data for such scenarios. Additionally, Spark includes the MLlib library that contains various machine learning algorithms.
 Though Spark has several advantages over Hadoop, it still leverages the Hadoop Distributed File System (HDFS) \cite{shvachko2010hdfs} technology for storage. 
 

 \textbf{Kibana:} Kibana is an open-source data visualization tool \cite{gupta2015kibana}. It is used for log data and time-series data analysis, as well as application monitoring. In addition, it provides enhanced dashboard capabilities. The data stored into an Elasticsearch index can be visualized using Kibana.

\section{Datasets}

In this section, we describe the datasets that we use to evaluate the performance of our model. Since our goal is to propose an intrusion detection system that can be utilized in big data scenarios, we primarily use a large-scale dataset collected from iSecurity that consists of real-world network traffic data (the iSecurity dataset contains 123 million instances whereas the total size of the dataset is above 1 TB).  However, the official implementations of most of the existing anomaly detection models \cite{ester1996dbscan,feng2006csoacn,liu2008isolation,choi2019unsupervisedautoencoder} are only evaluated on smaller-sized academic datasets while the available source codes of these models also did not handle big data scenarios. Thus, we additionally use various academic datasets to further validate the effectiveness of our approach by comparing our proposed model with the state-of-the-art anomaly detection models used in such datasets. In the following, we describe the datasets used in our experiments in detail.

\subsection{{Large Industrial Dataset}}

We used the log data of an iSecurity client from January 8th to January 14th as our training data. It contains 123,089,478 log instances, which is much larger than the size of the datasets used in previous work to train the Isolation Forest algorithm \cite{liu2008isolation}. For testing, we used the log data of the same client from January 15th to January 16th. An overall description of the dataset is shown in Table 1. Note that in the iSecurity dataset, there are about 100 attributes that are derived by Graylog's custom parser where most attributes are different metadata. After careful evaluation by the network experts at iSecurity, we select eight relevant attributes as features which are shown in Table 2. We can see from Table 2 that many of these attributes (e.g., timestamp, protocol names, IP addresses) contain non-numeric values. Thus, to give these attributes as input features to our proposed model, we convert them to numerical representations. Below, we describe our pre-processing technique.

\begin{table}
\centering
  \caption{An Overview of the iSecurity Dataset. `\#' denotes `total number of'.}
  \small
  \label{tab:isec_overview}
  \begin{tabular}{l|l|l}
    \toprule
    Type & '\#' Instances & Time Interval \\
    \midrule
     Training Data & 123,089,478 & January 8th 2019 (12:00 am) to January 14th 2019 (11:59 pm) \\
     Test Data & 34,184,556 & January 15th 2019 (12:00 am) to January 16th 2019 (11:59 pm) \\
  \bottomrule
  \end{tabular}
\end{table}

\begin{table*}
\centering
  \caption{Description of the Features in the iSecurity Dataset.}
  \small
  \label{tab:isec_features}
  \begin{tabular}{l|l}
    \toprule
    Attribute&Example\\
    \midrule
    Timestamp & January 14th 2019, 23:59:59.994 \\
        Host IP  & 192.168.213.12  \\
       Host Interface & inside, outside, Tunnel21(Site21), MNET, etc.  \\
 Target IP & 10.100.206.46\\ 
 Target Port & 443, 3389, etc. \\
 Target Interface & inside, outside, Tunnel32(Site32), F5WAF, etc. \\
 Protocol Name & TCP, UDP, ICMP, etc.  \\
 Action & Built, Permitted, Denied, etc. \\

  \bottomrule
  \end{tabular}
\end{table*}

\paragraph{\textbf{Timestamp:}} We pre-process the value of the timestamp $T\textsubscript{v}$ by converting it to the hour format via combining \textit{the day of the week} $D$ and \textit{the hour of the day} $H$ based on the following equation.
\begin{equation}
    T\textsubscript{v} = D * 24 + H
\end{equation}

Here, $D \in [0, 6]$ and $H \in [0, 23]$. We convert the timestamp value in this way in order to keep track of an incident happened in a specific hour on a particular day.
\paragraph{\textbf{IP Addresses (Host and Target):}} Since, all IP addresses may not be of equal size (e.g., 10.220.50.51, 192.168.1.1), we make all values before/after each dot of equal size (size = 3) by adding
leading zeroes. Then we remove all the dots and concatenate each byte of the IP address together as shown below: 
\begin{itemize}
    \item 10.220.50.51 $\rightarrow$ 010.220.050.051 $\rightarrow$ 010220050051
\end{itemize}

\paragraph{\textbf{Other Features (Interface, Protocol, Action):}} For other features such as the interface name, the protocol name, the action name, and etc., we first convert each of them to the lowercase format. Then we convert these features to numerical representations using StringIndexer\footnote{\url{https://spark.apache.org/docs/latest/ml-features.html\#stringindexer}}. An example to convert protocol names to numerical representations using this approach is shown below:

\begin{itemize}
   \item TCP $\rightarrow$ tcp $\rightarrow$ 10000
\end{itemize}



\subsection{Academic Datasets}

We further evaluate the effectiveness of our proposed IForest-KMeans model by comparing with other models on the following 12 academic datasets: (i) Http and (ii) Smtp datasets from KDD CUP 99 that are also used for network intrusion detection \cite{yamanishi2004linehttp}, (iii) Annthyroid, (iv) Arrhythmia, (v) Wisconsin Breast Cancer (Breastw), (vi) Forest Cover Type (ForestCover), (vii) Ionosphere, (viii) Pima, (ix) Satellite, (x) Shuttle \cite{asuncion2007uci}, (xi) Mammography, and (xii) Mulcross \cite{shi2006unsupervisedmulcross}. We select these datasets for evaluation since these datasets are widely used in recent years to evaluate various anomaly detection models \cite{liu2008isolation}. Moreover, another advantage of these datasets is that they include the ground truth labels of the anomaly instances and so they can be used to measure the performance of different models. Note that the anomaly detection models don't have access to the ground truth labels when they are trained and these labels are only used during the evaluation stage. Thus, the anomaly detection models are trained in an unsupervised way. Since the industrial dataset that we use in this paper only contains network traffic data, the selected academic datasets are also diverse in this regard as they contain data from various categories, such as disease, network, satellite, and etc. The overall statistics of these datasets are shown in Table \ref{tab:aca_overciew}. 


\begin{table*}
\centering
  \caption{An Overview of the Academic Datasets.}
  \small
  \label{tab:aca_overciew}
  \begin{tabular}{c|c|c|c}
    \toprule
    Dataset& Total Number of Instances& Number of Dimensions & Anomaly Rate \\
    \midrule
    Http (KDDCUP99)  & 567497 & 3 & 0.4\% \\
    ForestCover  & 286048 & 10 & 0.9\%\\
    Mulcross  & 262144 & 4 & 10\% \\
    Smtp (KDDCUP99) &  95156 & 3 & 0.003\% \\
    Shuttle &  49097 & 9 & 7\% \\
    Mammography &  11183 & 6 & 2\% \\
    Annthyroid &  6832 & 6 & 7\% \\
    Satellite &  6435 & 36 & 32\%\\
    Pima &  768 & 8 & 35\%\\
    Breastw &  683 & 9  & 35\%\\
    Arrhythmia  & 452 & 274 & 15\%\\
    Ionosphere &  351 & 32 & 36\% \\
  \bottomrule
  \end{tabular}
\end{table*}



\section{Implementation of IForest-KMeans}

In this section, we describe the implementation details of our proposed IForest-KMeans model. As mentioned earlier that our proposed IForest-KMeans model is trained in two steps: 
\begin{enumerate}
    \item First, we train the Isolation Forest model in the given dataset.
    \item Second, we train the K-Means model where the anomaly scores predicted by the Isolation Forest model are given as input. 
\end{enumerate}
 Since the MLlib library of Spark contains the official Spark-based implementation of K-Means, we use this library to implement the K-Means algorithm. However, it should be pointed out that Spark does not contain any official implementation of Isolation Forest. Thus, we utilize the Spark-based third-party library of Isolation Forest implemented by Yang et al. \cite{titicacasparkiforest}. Note that Yang et al. \cite{titicacasparkiforest} only used various small-sized datasets to evaluate this Isolation Forest library \cite{liu2008isolation} where the size of the evaluated datasets ranges from only 683 instances in the smallest-sized Wisconsin Breast Cancer Dataset \cite{asuncion2007uci} to the largest-sized $http$ \cite{yamanishi2004linehttp} dataset that contains 567,498 instances. In comparison to them, our training data in the iSecurity dataset contains more than 123 million instances (almost 217 times larger than the largest dataset used by Yang et al. \cite{titicacasparkiforest}). 
 
 When we train the Isolation Forest model using the library developed by Yang et al. \cite{liu2008isolation} in our iSecurity dataset, we observe that the existing library caused Memory Leaks\footnote{Java Out Of Memory Exception: \url{https://tinyurl.com/gosry4v}} (i.e., out of memory error) during the training process. While investigating the possible issues behind the memory leaks, we find that the existing library\footnote{\url{https://git.io/Jv9hL}}
used the $count()$\footnote{\url{https://spark.apache.org/docs/1.5.0/api/java/org/apache/spark/sql/DataFrame.html}} method to count the total number of training instances in the dataset. However, calling this method in Apache Spark is a very expensive approach in large datasets. We alleviate this issue by using a pre-determined constant value to set the total number of training instances instead of setting it via calling the $count()$ method. This simple modification allows the Isolation Forest model to overcome the memory leaks.

Moreover, we also observe that the Isolation Forest model is highly sensitive to the hyperparameters. Based on our experiments, we found that different parameter settings led the model to cause memory-related issues while training the model in large datasets. After some trials and errors, we found the best hyperparameters that could be used to train the Isolation Forest model in scenarios when the size of the dataset is very large. In the following, we first discuss the parameter settings of our models, followed by describing in detail how we train and evaluate the proposed IForest-KMeans model in large datasets stored in Elasticsearch. Finally, we describe how we extend our trained model for real-time detection of anomalies in industrial settings.

\subsection{Parameter Settings}

In this section, we describe the parameters that we use to train our IForest-KMeans model. For K-Means, we set the value of total numbers clusters $K = 2$. For Isolation Forest, the parameter settings which are set to train the model are described below.

\begin{itemize}
      \item \textbf{Number of Trees:} We set the value of the number of trees to 256, which was found effective in the original paper \cite{liu2008isolation}. 
    
    \item \textbf{Sub-sample size:} This is the value which determines the number of instances to be used in each isolation tree. We find that our model can be trained successfully with the value 1000 and any values larger than this was causing the out of memory problem. 
    
    \item \textbf{Contamination Ratio:} The value of the contamination ratio indicates the proportion of outliers in the data set. This value should be within the range (0, 1) where the default value is set to $0.1$. We also use the default value $0.1$ while training the Isolation Forest model. Note that this value is only used by the Isolation Forest during the prediction phase to classify the incidents by converting the predicted anomaly score of each incident to a predicted label \cite{titicacasparkiforest}. However, our proposed IForest-KMeans does not require the contamination ratio for prediction since it converts the anomaly scores to predicted labels using the K-Means algorithm. Furthermore, the Spark-based Isolation Forest library that we use in this paper also uses a threshold to convert the anomaly scores to different labels which is calculated by the Approximate Quantile \cite{titicacasparkiforest}.
    
    \item \textbf{Approximate Quantile Relative Error:} The Approximate Quantile is an important hyperparameter to train Isolation Forest algorithm in large datasets. It is used to get a threshold to convert the anomaly scores to the predicted labels. The relative error for Approximate Quantile should be within the range 0 to 1 (inclusive). When the value of relative error is 0, it will calculate the exact labels. However, the value of 0 would be very expensive for large datasets which could cause memory leaks \cite{titicacasparkiforest,linkedin}. We empirically select the value as $0.5$ to train Isolation Forest and found that any values less than $0.5$  would cause the out of memory error. Nevertheless, this value is also not required by the IForest-KMeans model during the prediction phase to convert the anomaly scores to different labels. This is due to the fact that our proposed IForest-KMeans model utilizes K-Means instead of using approximate quantile to convert the predicted scores to different labels. 
\end{itemize}


\subsection{Anomaly Detection on Large Network Traffic Data in Elasticsearch}

Similar to the utilization of Python Programming Language in most prior work to implement different anomaly detection algorithms \cite{domingues2018comparative}, we also use the Python API of Spark: PySpark\footnote{\url{https://spark.apache.org/docs/latest/api/python/pyspark.html}}. The training and evaluation algorithm of our IForest-KMeans model for anomaly detection is shown in Algorithm 1. Below, we describe our algorithm. 


\begin{algorithm*}
\small
\raggedright{\textbf{Data:} Historical data from elasticsearch or Real-time data from live streams}\\
\raggedright{\textbf{Type:} Training or Evaluation}\\
\raggedright{\textbf{IForestKMeans(Data, Type):}}
\begin{algorithmic}
\STATE $RDD\textsubscript{} \leftarrow createRDD(Data)$ 
\STATE $DF \leftarrow createDataFrame(RDD\textsubscript{})$  
\STATE $DF\textsubscript{\textit{P}} \leftarrow PreprocessDataFrame(DF)$ 
\IF{$Type = Training$}
\STATE  $Model\textsubscript{\textit{IForest}} \leftarrow IsolationForest.Train(DF\textsubscript{\textit{P}})$ 
\STATE $DF\textsubscript{\textit{AS}} \leftarrow getAnomalyScore(Model\textsubscript{\textit{IForest}})$
\STATE  $Model\textsubscript{\textit{KMeans}} \leftarrow KMeans.Train(DF\textsubscript{\textit{AS}})$ 

\STATE  $Model\textsubscript{\textit{IForestKMeans}} \leftarrow [Model\textsubscript{\textit{IForest}}, Model\textsubscript{\textit{KMeans}}]$ 

\STATE  $Model\textsubscript{\textit{IForestKMeans}}.Save(Path)$  \COMMENT{Saving the trained models in the local disk}
\ELSE 
\STATE $Model\textsubscript{\textit{IForestKMeans}}\leftarrow Model.Load(Path)$ 
\STATE  $DF\textsubscript{\textit{Result}}  \leftarrow Model\textsubscript{\textit{IForestKMeans}}.Predict(DF\textsubscript{\textit{P}})$ 
\STATE  $ElasticsearchIndex.Write(DF\textsubscript{\textit{Result}})$ \COMMENT{Saving the result in Elasticsearch}
\ENDIF
\end{algorithmic}
\caption{IForest-KMeans for Network Intrusion Detection}
\end{algorithm*}

Since in big data scenarios, Spark utilizes the concept of Resilient Distributed Datasets (RDD) to operate data in parallel \cite{zaharia2016apachespark}, we first load all the training data from the Elasticsearch index into the Spark RDD. Due to the advantage of Spark's DataFrame over RDD while using different machine learning libraries of Spark \cite{meng2016mllib}, we converted the loaded RDD into DataFrame. Note that the Spark DataFrame is a dataset which is conceptually equivalent to a table in a relational database. It organizes each attribute in the dataset into named columns. We then pre-process the data in each column of the DataFrame based on our approach given in Section 6.1 followed by combining the pre-processed data of each column into a single feature vector via utilizing the Vector Assembler\footnote{\url{https://spark.apache.org/docs/latest/ml-features}} method.
This feature vector is used as the features which we give as input to the Isolation Forest model. After we finish training the Isolation Forest model, we extract the anomaly score of each instance predicted by the model. The anomaly scores are then used as features to train the K-Means model  to partition each instance into $K = 2$ clusters. Then we save the trained model into the local disk. Since anomalies are usually fewer than normal instances, we consider the cluster with more instances as the normal one while considering the cluster with fewer instances as the anomalous cluster. Note that most of the instances in our training data are also normal traffic patterns.

For evaluation, we load the trained IForest-KMeans model from the local disk. The loaded model first predict the anomaly score for each instance in the test set based on the trained Isolation Forest model. Then the predicted anomaly scores of all instances are given as input to the trained K-Means model in order to assign each data point to one of the clusters. Finally, we save the predicted result in our machine learning index of Elasticsearch and visualize it using Kibana to analyze the performance of our model.

\subsection{Real-time Anomaly Detection on Streaming Data}

We also evaluate our trained IForest-KMeans model on real-time network traffic data. For that purpose, we utilize the Spark Streaming \cite{zaharia2016apachespark}. The Spark Streaming is an extension of Spark API that enables processing of live data streams with scalability, high-throughput, and fault-tolerance \cite{zaharia2016apachespark}. The live data streams which are received by Spark Streaming are divided into batches. In Spark Streaming, data can be ingested from different sources which can be processed using different machine learning models. The processed data can also be stored in the file-systems or databases. Moreover, Spark utilizes its Structured Streaming \cite{armbrust2018structured} engine to process the streaming data.  

To evaluate our proposed model for real-time detection of anomalies in streaming data, we send the test data as live streams instead of reading them from Elasticsearch. Then Spark utilizes its Structured Streaming engine to process these data in real-time for anomaly detection. The results predicted by our trained model in the live streams are then stored into an Elasticsearch index for further analysis. It is to be noted that we do this experiment on iSecurity's server to investigate how our trained model performs in an industrial setup on real-time. 



\section{Result and Discussions}



In this section, we discuss the performance of our proposed IForest-KMeans model in a large industrial dataset as well as in academic datasets. We use the Isolation Forest as the baseline since this algorithm provided state-of-the-art performance on various unsupervised anomaly detection tasks \cite{chandola2009anomalysurvey,ahmed2016anomalysurvey}. Below, we first demonstrate the performance of our model on real-world network traffic data at iSecurity followed by discussing the performance on academic datasets.



\subsection{Performance on Real World Network Traffic Data at iSecurity}

In the original Isolation Forest algorithm, the trained model learns to predict the anomaly score of each data point. The higher the anomaly score, the higher the probability of an instance being an anomaly. In contrast, an anomaly score closer to 0 means it has a higher chance to be a normal instance. Note that the Spark-based implementation of the Isolation Forest \cite{titicacasparkiforest} uses the Approximate Quantile Relative Error (AQRE) to convert the predicted anomaly scores to different labels. However, as mentioned in Section 6.2.1, setting the value of AQRE to 0 to calculate the optimal labels is a very expensive process that also leads to memory leaks. To avoid the memory leaks, we increased the AQRE value to 0.5 while training the Isolation Forest model. In addition, we use another Isolation Forest baseline that uses the default AQRE value = 0 while utilizing only 1\% of the training set\footnote{The 1\% data were selected from the first 2 days of the training set since larger datasets with AQRE = 0 led to memory leaks.}. Similar to storing the predicted result of our proposed IForest-KMeans into an Elasticsearch index, the predicted results of the baselines are also indexed into an Elasticsearch index. Afterwards, we visualize the predicted results of all models using Kibana for analysis.

In the following, we compare the performance of our proposed IForest-KMeans model with the baselines in the iSecurity dataset based on two different use-cases. Since we don't have a true label for each event in the iSecurity dataset except for the instances used in our use-cases given below, we don't evaluate the performance for these use-cases using any evaluation metrics. Rather, we analyze the performance via visualizing with Kibana (see Figure \ref{fig:SigAnPred3389} and Figure \ref{fig:SigAnPredIP}).


\subsubsection{Use Case 1: Cyberattack on a Target Port}

\begin{figure*}[t!]
\begin{center}
\includegraphics[width=\linewidth]{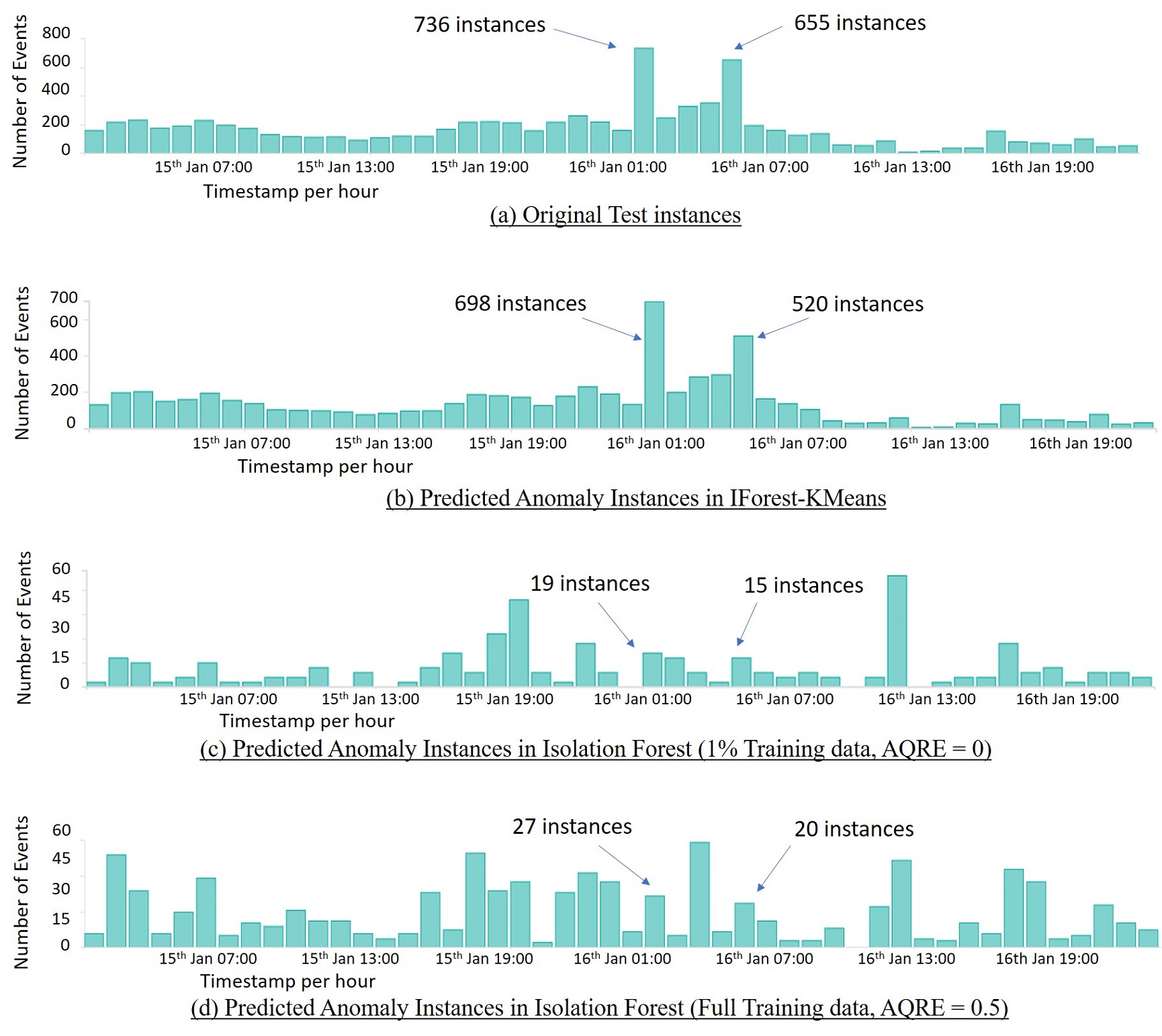}

\caption{
Performance on detecting cyberattacks on the Target Port 3389. Using Kibana, we visualize the (a) Original Test Instances during the time interval: January 15th to January 16th, (b) Instances predicted as anomaly by the IForest-KMeans model during that time interval, (c) Instances predicted as anomaly by the \textit{Isolation Forest (1\% training data, AQRE = 0)} model during that time interval, and (d) Instances predicted as anomaly by the \textit{Isolation Forest (Full training data, AQRE = 0.5)} model during that time interval.
}
\label{fig:SigAnPred3389}
\end{center}
\end{figure*}

In this use-case, we demonstrate the performance of our model in scenarios when a cybarattack occurs by targeting a specific port in the iSecurity dataset (as illustrated in Section 3). In Figure \ref{fig:SigAnPred3389}, we show the result predicted as anomalies by different models for the cyberattacks on target port 3389. We observe from Figure \ref{fig:SigAnPred3389} that the IForest-KMeans detects the anomalies much better than the baselines that are based on \textit{Isolation Forest}. We find that during the time interval (16th January, from 1 am to 6 am) when the cyberattacks targeted the port 3389, the IForest-KMeans predicted about 6,693 instances as potential anomalies among 8,296 instances in the test data. On the other hand, the \textit{Isolation Forest (AQRE = 0.5)} predicted only 965 instances as anomalies while the \textit{Isolation Forest (AQRE = 0 and utilized only 1\% training data)} predicted only 720 instances as anomalies. Moreover, around 1 am and 5 am when most of the attacks occurred, the IForest-KMeans predicted 698 instances among 736 instances as anomalies for the time interval 1 am to 2 am, and predicted 520 instances among 655 instances as anomalies for the time interval 5 am to 6 am. However, both baselines based on \textit{Isolation Forest} model performed very poorly in these cases. The \textit{Isolation Forest (AQRE = 0.5)} only predicted 27 instances as anomalies for the time interval 1-2 am and 20 instances for the time interval 5-6 am. In case of the other Isolation Forest baseline that utilized only 1\% of the training data, 19 instances were predicted as anomalies for the time interval 1-2 am while 15 instances were predicted as anomalies for the time interval 5-6 am. 
\subsubsection{Use Case 2: Cyberattack from a Specific IP}

\begin{figure*}[t!]
\begin{center}
\includegraphics[width=\linewidth]{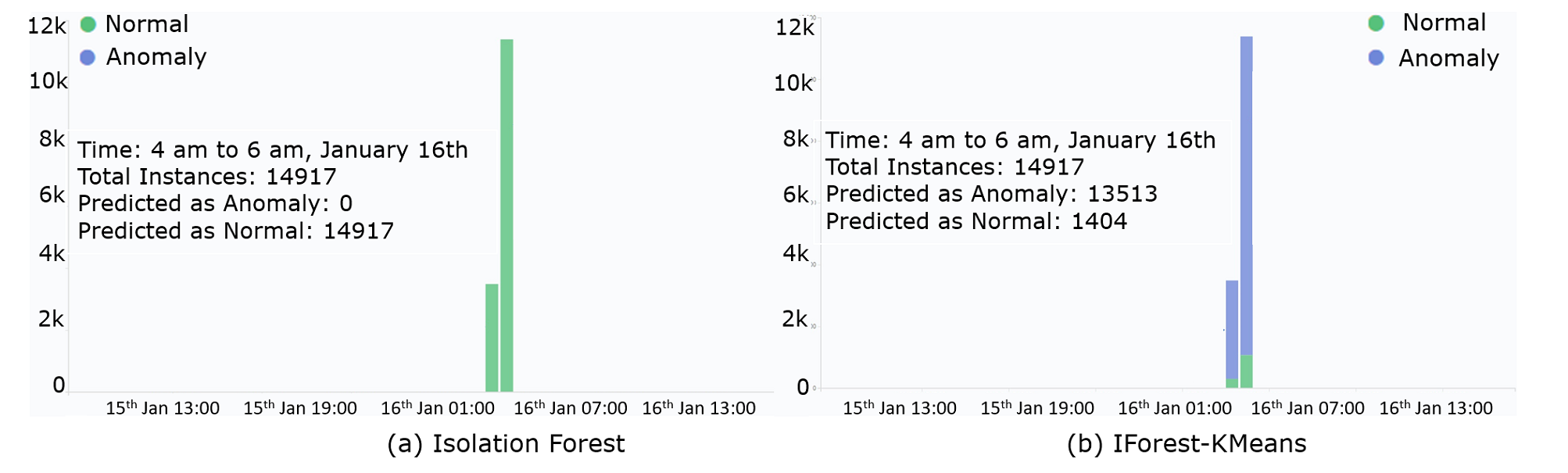}

\caption{
The performance of different models on detecting cyberattacks from the IP Address 10.100.208.115 during the time interval: January 15th to January 16th. Based on Kibana Visualization: (a) Prediction by the \textit{Isolation Forest} baselines and (b) Prediction by the \textit{IForest-KMeans} model during that time interval. Here, X-axis represents timestamp and Y-axis represents total number of events. 
}
\label{fig:SigAnPredIP}
\end{center}
\end{figure*}

For the case when the cyberattack occurred from the IP address 10.100.208.115 (as illustrated in Section 3), we show the predicted result of our proposed model as well as the baseline Isolation Forest model\footnote{For this use-case, since we observe the same result for both baselines, we show them together in Figure \ref{fig:SigAnPredIP}.} in Figure \ref{fig:SigAnPredIP}. While comparing our IForest-KMeans model with the baseline \textit{Isolation Forest} model, we find that the IForest-KMeans again outperforms the baseline. However, this time the performance of the \textit{Isolation Forest} model in cases of both \textit{AQRE = 0.5} and \textit{AQRE = 0 while using 1\% of the training data} is even worse as they fail to detect a single instance as anomalous. Among the 14,917 instances (From January 15, 12 am to January 16, 11:59 pm) that occurred by the host IP 10.100.208.115, our proposed IForest-KMeans model detected 13,513 instances as anomalies. We find that during the time interval (16th January, from 4 am to 5 am) when the IP address 10.100.208.115 was the host of 3,492 traffic instances, the IForest-KMeans model predicted about 3,188 instances as potential anomalies among them. Moreover, within the time interval 5 am to 6 am, when most of the attacks occurred from that host IP 10.100.208.115, the IForest-KMeans detected 10,322 instances as anomalies among the 11,396 instances. 
If we consider all 14,917 instances that occurred by the host IP 10.100.208.115 within this time-interval (January 15th to January 16th) as anomalies, then the overall accuracy of IForest-KMeans is 91\%. On the contrary, the baseline Isolation Forest could not detect any anomalous instances during that time-interval. Since we observe that the Spark-based Isolation Forest model cannot be trained on large datasets without increasing the $AQRE$ value, we also demonstrate that setting the value of $AQRE = 5$ to train the \textit{Isolation Forest} model causes huge performance deterioration. Moreover, we also demonstrate that while using $AQRE = 0$, we could only train using 1\% of the total training data that also leads to huge performance degradation. The impressive detection rate of our proposed model in comparison to the \textit{Isolation Forest} baselines shows that our model can be utilized in large datasets without compromising the performance.

\subsubsection{Performance on Streaming Data} We utilize Spark's Structured Streaming engine to investigate the effectiveness of our trained model on the live streaming data in iSecurity's system architecture for network intrusion detection. Based on our experiments on live data steams, we observe that our proposed network intrusion detection system can process the data stream in real-time along with instantly storing the predicted result into an Elasticsearch Index (it took about $0.65$ Millisecond on average to process each instance of the log data). This indicates that our proposed model can be effectively utilized in a real-world large-scale industrial domain for the real-time detection of anomalies in live data streams. 

\subsubsection{Time Complexity Analysis} The Spark's MLlib implementation of the K-Means algorithm uses the parallel variant of the K-Means++ \cite{arthur2007kmeans++} method called K-Means|| \cite{bahmani2012scalablekmeans}. Note that K-Means++ \cite{arthur2007kmeans++,bachem2016fastkmeansnips,franti2019howmuchkmeans} is an extension of the original K-Means algorithm that improves the speed and the accuracy of the original method; while the K-Means|| parallelizes the K-Means++ method that also allowed it to be effectively utilized in large datasets \cite{bahmani2012scalablekmeans,hamalainen2021improvingscalablekmeans,franti2019howmuchkmeans}. Given a dataset of size = $n$, the number of clusters = $k$, and the input dimension = $d$, the time complexity of K-Means|| is $\mathcal{O}(nkd\log{}n)$ \cite{bachem2016fastkmeansnips}. Since the Isolation Forest is a linear algorithm \cite{liu2008isolation}, for datasets having $n$ instances, the time complexity of Isolation Forest is $\mathcal{O}(n)$. Thus, when the KMeans algorithm is trained using the anomaly scores predicted by the Isolation Forest, the overall time complexity becomes $\mathcal{O}(n)$ $+$ $\mathcal{O}(nkd\log{}n)$. Since  the input dimension of the anomaly scores in IForest-KMeans is 1, and due to using 2 clusters (normal, anomaly) for the iSecurity dataset, the value of $k$ becomes 2 and the value of $d$ becomes 1. Excluding these constant values, the overall time complexity of our proposed IForest-KMeans can be written as: 
\begin{equation}
    \mathcal{O}(n) + \mathcal{O}(n\log{}n) \approx \mathcal{O}(n\log{}n) 
\end{equation}

Note that we ran our experiments on an iSecurity server machine that had an Intel Xeon Processor with 32 cores and 128 GB of Memory. It took about one week to train the IForest-KMeans model on the whole training dataset of iSecurity. However, once the model is trained, it could effectively be utilized (only about $0.65$ Millisecond on average to process each instance of the log data) for anomaly detection on real-time streaming data, as illustrated in section $8.1.3$. Moreover, since our proposed model is based on machine learning techniques, it can also be periodically updated offline by re-training on new training data. Thus, once the update is done, the newly trained model can be used to replace the online version. In this way, the model will be able to capture the pattern changes in the live streaming data. It is to be noted that such a way of updating a model is also a very common practice in industries \cite{gama2014surveyconceptdrift}. 
\subsection{Performance on Academic Datasets}

\begin{table*}
  \caption{Performance on Academic Datasets in terms of AUC score. Here, `$\dagger$', denotes the Spark-based implementation of Isolation Forest by Yang et al. (\cite{titicacasparkiforest}) where the result is obtained by us, while `$\S$' denotes the result of the original implementation of the Isolation Forest algorithm obtained by Liu et al. \cite{liu2008isolation}.}
  \centering
  \label{tab:aca_result}
  \small
  \begin{tabular}{p{3cm}|p{1.25cm}|p{1.25cm}|p{1.25cm}|p{1.25cm}|p{1.25cm}|p{1.25cm}}
    \toprule
    Dataset & IForest-KMeans & Isolation Forest `$\dagger$' \cite{titicacasparkiforest} & Isolation Forest `$\S$' \cite{liu2008isolation} & ORCA \cite{liu2008isolation} & LOF \cite{liu2008isolation} & RF \cite{liu2008isolation}\\
    \midrule
    Http (KDD CUP 99)  & 0.96 & 0.97 & 1.00 & 0.36 & N/A & N/A  \\
    ForestCover  & 0.88 & 0.78 & 0.88 & 0.83 & N/A & N/A \\
    Mulcross  & 0.92 & 0.85 & 0.97 & 0.33 & N/A & N/A \\
    Smtp (KDD CUP 99) &  0.85 & 0.86 & 0.88 & 0.80 & N/A & N/A \\
    Shuttle &  0.98 & 0.97  & 1.00 & 0.60 & 0.55 & N/A \\
    Mammography &  0.82 & 0.76 & 0.86 & 0.77 & 0.67 & N/A \\
    Annthyroid &  0.75 & 0.66 & 0.82 & 0.68 & 0.72 & N/A \\
    Satellite & 0.73 & 0.67 & 0.71 & 0.65 & 0.52 & N/A \\
    Pima &  0.64 & 0.58 & 0.67 & 0.71 & 0.49 & 0.65 \\
    Breastw &  0.98 & 0.67  & 0.99 & 0.98 & 0.37 & 0.97 \\ 
    Arrhythmia  & 0.76 & 0.65 & 0.80 & 0.78 & 0.73 & 0.60 \\
    Ionosphere & 0.78 & 0.66 & 0.85 & 0.92 & 0.89 & 0.85 \\
  \bottomrule
  \end{tabular}
\end{table*}

For the academic datasets, we compare the performance of our model with the following models: the Spark-based implementation of Isolation Forest \cite{titicacasparkiforest}, the original Isolation Forest algorithm \cite{liu2008isolation,liu2012isolation}, ORCA \cite{orca}, LOF \cite{lof}, and Random Forest (RF) \cite{shi2006unsupervisedmulcross}. Note that we obtain the results for ORCA, LOF, and RF from Liu et al. \cite{liu2008isolation}. For the PySpark-based Isolation Forest, we use the default parameters that were used in the original paper \cite{liu2008isolation} to run the experiments. In Table \ref{tab:aca_result}, we show the results of different models on the publicly available datasets commonly used in academia. Similar to the prior work, we compare the results based on the Area Under Curve (AUC) metric \cite{myerson2001areaAUC1,pruessner2003twoAUC2}.

We find from Table \ref{tab:aca_result} that our proposed IForest-KMeans outperforms the Spark-based implementation of Isolation Forest: PySpark Isolation Forest \cite{titicacasparkiforest} in most datasets. Among the 12 datasets, our proposed IForest-KMeans outperforms the PySpark Isolation Forest in 10 of them. The PySpark Isolation Forest only performs better than our model in two datasets (Http and Smtp). Based on paired t-test ($p$ $\leq$ $.05$), the performance difference between the IForest-KMeans and PySpark Isolation Forest is \textbf{statistically significant}. Furthermore, we find that our model's results obtained in these datasets are more comparable with the result obtained by Liu et al. in their original implementation of Isolation Forest \cite{liu2008isolation}. We also find based on paired t-test ($p$ $\leq$ $.05$) that the performance of our model is \textbf{significantly} better than ORCA, LOF, and RF in most of the datasets \cite{liu2008isolation,orca,lof,shi2006unsupervisedmulcross}. 

While comparing the results mentioned in the original paper \cite{liu2008isolation} with the results obtained by us using the PySpark Isolation Forest \cite{titicacasparkiforest}, we find that the results are not reproducible. Note that the result of the original paper was also not reproduced in many datasets by Yang et al. \cite{liu2008isolation} when they compared the result of the PySpark Isolation Forest with the original model. It should be pointed out that the source code of the original Isolation Forest model was not shared by Liu et al. \cite{liu2008isolation} who proposed this model and the original model was only proposed for non-distributed settings.

The other advantage of our proposed model is that it requires fewer parameters than different Spark-based implementations of Isolation Forest \cite{titicacasparkiforest,linkedin}. For instance, our model does not require the following parameters: the contamination rate, and the approximate quantile relative error, that are required by the PySpark Isolation Forest \cite{titicacasparkiforest}. However, without leveraging these parameters, our proposed model outperforms the PySpark Isolation Forest in most datasets. Below, we conduct some case studies to further analyze the effectiveness of the IForest-KMeans model.





\subsubsection{Case Study: Classification Performance}

\begin{figure*}[t!]
\begin{center}
\includegraphics[width=\linewidth]{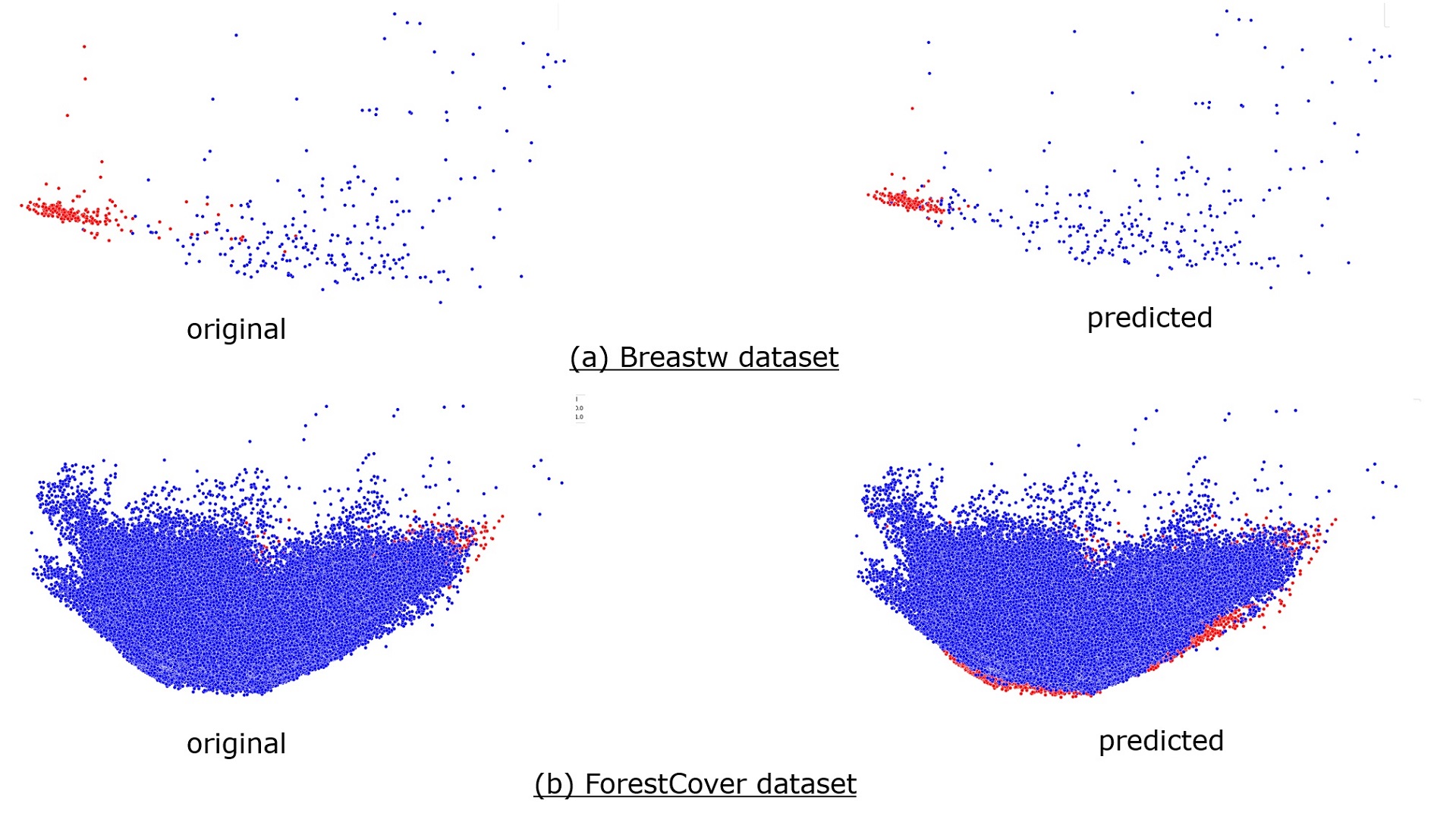}

\caption{
Classification performance of the proposed IForest-KMeans model based on PCA. `Red' color denotes `Anomaly' instances, while `Blue' color denotes `Normal' instances.
}
\label{fig:academicanalysis}
\end{center}
\end{figure*}

In this section, we analyze the classification performance of the IForest-KMeans model. For that purpose, we use the Breastw and ForestCover datasets and apply dimensionality reduction via utilizing the Principal Component Analysis (PCA) \cite{wold1987principal} to project the data into a two-dimensional space for analysis as shown in Figure \ref{fig:academicanalysis}. We find from Figure \ref{fig:academicanalysis} that the proposed IForest-KMeans model was effective to classify most of the anomaly instances in both datasets. While the proposed IForest-KMeans model also had some false positives in case of the ForestCover dataset, the number of false positives is still quite low since it could correctly classify most of the normal instances.

\subsubsection{Case Study: Hyperparameter Sensitivity}

\begin{figure*}[t!]
\begin{center}
\includegraphics[width=\linewidth]{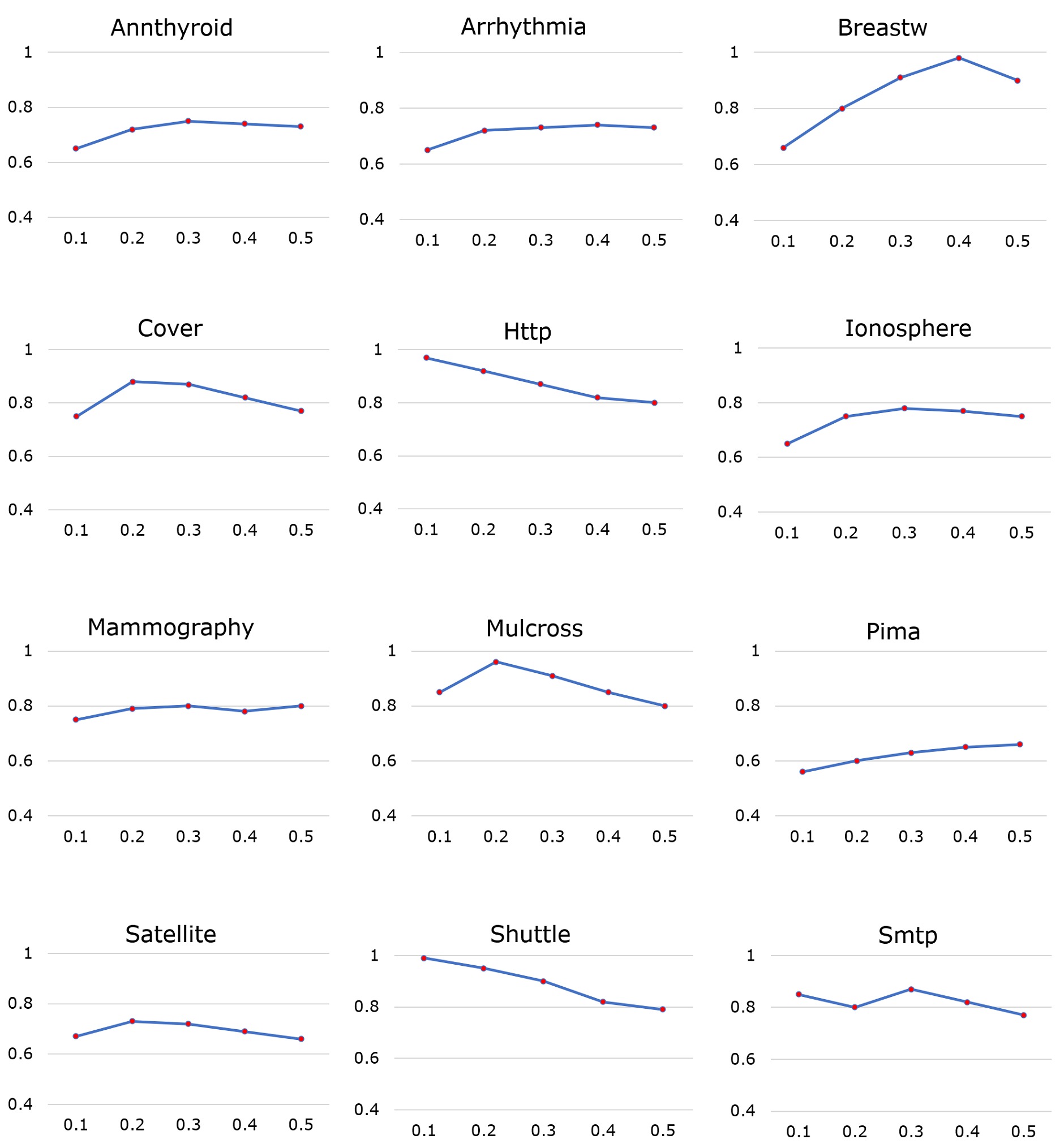}

\caption{
Performance of Isolation Forest in each academic dataset based on various contamination ratio $[0.1, 0.5]$. For each dataset, the X-axis represents the contamination ratio and the Y-axis represents the AUC score.
}
\label{fig:mergecr}
\end{center}
\end{figure*}

\begin{figure*}[t!]
\begin{center}
\includegraphics[width=\linewidth]{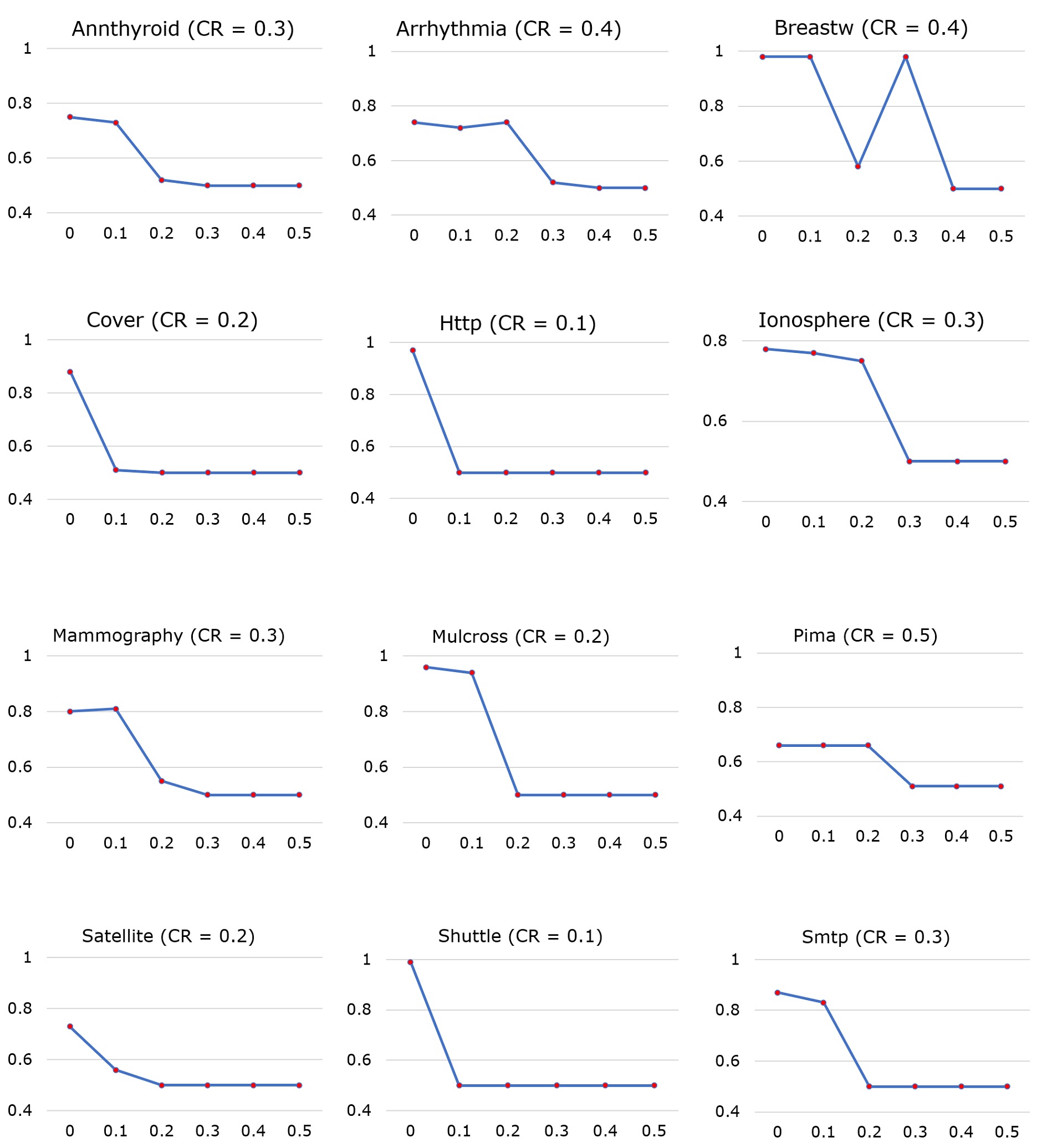}

\caption{
Performance of Isolation Forest in each academic dataset based on the optimal Contamination Ratio (CR) with various values of AQRE $[0, 0.5]$.  For each dataset, the X-axis represents the value of AQRE and the Y-axis represents the AUC score.
}
\label{fig:mergeaqre}
\end{center}
\end{figure*}

Here, we investigate the sensitivity of following hyperparameters in Spark-based Isolation Forest \cite{titicacasparkiforest,linkedin} which are not required for the IForest-KMeans model: 
\begin{itemize}
    \item Contamination Ratio (CR): It indicates the ratio of anomaly in the dataset. This is used to convert the predicted anomaly scores to the labels. 
    \item Approximate Quantile Relative Error (AQRE): As discussed in Section 7.1, this is the relative error used for the approximate quantile calculation to assign labels to the anomaly scores. 
\end{itemize}

Note that on the PySpark Isolation Forest library \cite{titicacasparkiforest} that we use in this paper, the default value of CR is set to 0.1 while the default AQRE value is set to 0. We show the result of Isolation Forest in each academic dataset by varying the value of CR (within the range $0.1$ to $0.5$, inclusive) in Figure \ref{fig:mergecr}. From Figure \ref{fig:mergecr}, we find that different values of CR give different results in each dataset. Specially, in the Breastw dataset, we find that the AUC score is increased while the value of CR is increased (up-to the CR value 0.4). More specifically, with increasing the value of CR from the default value 0.1 to the new value 0.4, the performance is increased by 48\%. However, in the Http dataset, the increasing value of CR decreases the performance. Thus, it indicates that the performance of the Isolation Forest model is sensitive to the value of CR. Moreover, we find that the difference in performance when the default value of CR is used and when the optimal value of CR is used is \textbf{statistically significant} based on paired t-test ($p$ $\leq$ $.05$).  It should be noted that the real-world datasets are usually unlabeled, and so such datasets don't contain information about the ratio of anomaly. In contrary to the Isolation Forest algorithm, our proposed IForest-KMeans does not require the information of CR but still provides performance almost similar to the Isolation Forest.

Moreover, in Figure \ref{fig:mergeaqre}, we show the effect of another hyperparameter AQRE in the Isolation Forest model. Here, for each dataset, we use the optimal value of CR and change the AQRE value within the range $0$ to $0.5$ (inclusive). We find that in most datasets, the performance is decreased while we increase the value of AQRE. Thus, it implies that to predict more accurate labels, the Isolation Forest algorithm requires lower values of AQRE. Recall that we set the AQRE value to $0.5$ while training the large iSecurity dataset since we observe that the Isolation Forest algorithm cannot be trained on large datasets if the AQRE value is lower than 0.5. However, we find that setting the AQRE value to $0.5$ decreases the performance in all 12 academic datasets. Based on paired t-test ($p$ $\leq$ $.05$), the performance deterioration in Isolation Forest when the default value of $0.1$ is used for AQRE and when the AQRE value is set to $0.5$ is \textbf{statistically significant}. Our extensive experiments demonstrate that even after using fewer parameters, the proposed IForest-KMeans model is still effective in both large and small datasets.

\section{Conclusions and Future Work}

 In this paper, we propose an intrusion detection system to detect anomalies in large network traffic data. First, we address the issues that prevent the Isolation Forest algorithm from being trained on large industrial datasets. Then we demonstrate how we overcome these limitations for the future replicability of experiments in such large datasets. For our proposed intrusion detection system, we introduce a novel machine learning approach via combining Isolation Forest with K-Means that also alleviates the issues to train the Isolation Forest algorithm in big data scenarios. Based on our experiments in real-world large network traffic data at iSecurity, we find that our proposed intrusion detection system can be effectively trained on a large industrial dataset while using fewer parameters than the original Isolation Forest model to provide impressive performance for anomaly detection. Moreover, we compare the effectiveness of our proposed approach with other state-of-the-art models on 12 academic datasets. Finally, the usefulness of our proposed intrusion detection system in the industrial domain is also verified as we show that our system is effective for real-time detection of anomalies on the live streaming data at iSecurity. 
 
  In the future, we will evaluate our proposed system in more use-cases along with investigating how deep learning and other machine learning-based models perform on anomaly detection in big data scenarios on other cyber-physical systems \cite{tcps1,tcps2,tcps3,tcps4,tcps5,tcps6,tcps7,tcps8}. Moreover, we will also investigate machine learning algorithms in industrial big data scenarios on other domains \cite{zobaed2019clustcrypt,munir2019localized,hossain2019distortion,newaz2019healthguard}, such as Information Retrieval \cite{JH1,JH2,JH3,JH5,JH6,JH4,JH10}, Natural Language Processing \cite{vaswani2017attention,devlin2019bert,laskar-LREC,laskar2020query,bari2019zero,laskar2020utilizing,bari2020multimix,laskar2020wslds}, and etc. 

\section*{Acknowledgements}
We gratefully appreciate the associate editor and the reviewers for their excellent review comments. We thank Khalid Chatha, Jan Jedrasik, and many other employees at Dapasoft and iSecurity who helped us by preparing the large iSecurity dataset, setting up the computing servers for experiments, along with conducting the evaluation. This research is supported by a Discovery Grant from the Natural Sciences and Engineering Research Council of Canada, an Ontario Research Fund-Research Excellence award in BRAIN Alliance, and the York Research Chairs (YRC) program.

\bibliographystyle{plain}  
\bibliography{references}

\begin{thebibliography}{10}

\bibitem{ahmed2016anomalysurvey}
Mohiuddin Ahmed, Abdun~Naser Mahmood, and Jiankun Hu.
\newblock A survey of network anomaly detection techniques.
\newblock {\em Journal of Network and Computer Applications}, 60:19--31, 2016.

\bibitem{aloqaily2019intrusiondecision}
Moayad Aloqaily, Safa Otoum, Ismaeel Al~Ridhawi, and Yaser Jararweh.
\newblock An intrusion detection system for connected vehicles in smart cities.
\newblock {\em Ad Hoc Networks}, 90:101842, 2019.

\bibitem{tcps1}
Riham Altawy and Amr~M Youssef.
\newblock Security, privacy, and safety aspects of civilian drones: A survey.
\newblock {\em ACM Transactions on Cyber-Physical Systems}, 1(2):1--25, 2016.

\bibitem{amer2013enhancingunsupervisedSVM}
Mennatallah Amer, Markus Goldstein, and Slim Abdennadher.
\newblock Enhancing one-class support vector machines for unsupervised anomaly
  detection.
\newblock In {\em Proceedings of the ACM SIGKDD Workshop on Outlier Detection
  and Description}, pages 8--15, 2013.

\bibitem{armbrust2018structured}
Michael Armbrust, Tathagata Das, Joseph Torres, Burak Yavuz, Shixiong Zhu,
  Reynold Xin, Ali Ghodsi, Ion Stoica, and Matei Zaharia.
\newblock Structured streaming: A declarative api for real-time applications in
  apache spark.
\newblock In {\em Proceedings of the 2018 International Conference on
  Management of Data}, pages 601--613, 2018.

\bibitem{arthur2007kmeans++}
David Arthur and Sergei Vassilvitskii.
\newblock k-means++ the advantages of careful seeding.
\newblock In {\em Proceedings of the eighteenth annual ACM-SIAM symposium on
  Discrete algorithms}, pages 1027--1035, 2007.

\bibitem{asuncion2007uci}
Arthur Asuncion and David Newman.
\newblock Uci machine learning repository, 2007.

\bibitem{bachem2016fastkmeansnips}
Olivier Bachem, Mario Lucic, S~Hamed Hassani, and Andreas Krause.
\newblock Fast and provably good seedings for k-means.
\newblock In {\em Proceedings of the 30th International Conference on Neural
  Information Processing Systems}, pages 55--63, 2016.

\bibitem{bahmani2012scalablekmeans}
Bahman Bahmani, Benjamin Moseley, Andrea Vattani, Ravi Kumar, and Sergei
  Vassilvitskii.
\newblock Scalable k-means+.
\newblock {\em Proceedings of the VLDB Endowment}, 5(7), 2012.

\bibitem{bari2019zero}
M~Saiful Bari, Shafiq Joty, and Prathyusha Jwalapuram.
\newblock {Zero-Resource Cross-Lingual Named Entity Recognition}.
\newblock {\em arXiv preprint arXiv:1911.09812}, 2019.

\bibitem{bari2020multimix}
M~Saiful Bari, Muhammad~Tasnim Mohiuddin, and Shafiq Joty.
\newblock Multimix: A robust data augmentation strategy for cross-lingual nlp.
\newblock {\em arXiv preprint arXiv:2004.13240}, 2020.

\bibitem{orca}
Stephen~D Bay and Mark Schwabacher.
\newblock Mining distance-based outliers in near linear time with randomization
  and a simple pruning rule.
\newblock In {\em Proceedings of the ninth ACM SIGKDD international conference
  on Knowledge discovery and data mining}, pages 29--38, 2003.

\bibitem{tcps6}
Lei Bu, Wen Xiong, Chieh-Jan~Mike Liang, Shi Han, Dongmei Zhang, Shan Lin, and
  Xuandong Li.
\newblock Systematically ensuring the confidence of real-time home automation
  iot systems.
\newblock {\em ACM Transactions on Cyber-Physical Systems}, 2(3):1--23, 2018.

\bibitem{chalapathy2019deepsurvey}
Raghavendra Chalapathy and Sanjay Chawla.
\newblock Deep learning for anomaly detection: A survey.
\newblock {\em arXiv preprint arXiv:1901.03407}, 2019.

\bibitem{chandola2009anomalysurvey}
Varun Chandola, Arindam Banerjee, and Vipin Kumar.
\newblock Anomaly detection: A survey.
\newblock {\em ACM computing surveys (CSUR)}, 41(3):1--58, 2009.

\bibitem{tcps8}
Sujit~Rokka Chhetri, Arquimedes Canedo, and Mohammad Abdullah~Al Faruque.
\newblock Confidentiality breach through acoustic side-channel in
  cyber-physical additive manufacturing systems.
\newblock {\em ACM Transactions on Cyber-Physical Systems}, 2(1):1--25, 2017.

\bibitem{choi2019unsupervisedautoencoder}
Hyunseung Choi, Mintae Kim, Gyubok Lee, and Wooju Kim.
\newblock Unsupervised learning approach for network intrusion detection system
  using autoencoders.
\newblock {\em The Journal of Supercomputing}, 75(9):5597--5621, 2019.

\bibitem{corsini2006combining}
Paolo Corsini, Beatrice Lazzerini, and Francesco Marcelloni.
\newblock Combining supervised and unsupervised learning for data clustering.
\newblock {\em Neural Computing \& Applications}, 15(3-4):289--297, 2006.

\bibitem{dean2010mapreduce}
Jeffrey Dean and Sanjay Ghemawat.
\newblock Mapreduce: a flexible data processing tool.
\newblock {\em Communications of the ACM}, 53(1):72--77, 2010.

\bibitem{devlin2019bert}
Jacob Devlin, Ming-Wei Chang, Kenton Lee, and Kristina Toutanova.
\newblock Bert: Pre-training of deep bidirectional transformers for language
  understanding.
\newblock In {\em Proceedings of the 2019 Conference of the North American
  Chapter of the Association for Computational Linguistics: Human Language
  Technologies, Volume 1 (Long and Short Papers)}, pages 4171--4186, 2019.

\bibitem{domingues2018comparative}
R{\'e}mi Domingues, Maurizio Filippone, Pietro Michiardi, and Jihane Zouaoui.
\newblock A comparative evaluation of outlier detection algorithms: Experiments
  and analyses.
\newblock {\em Pattern Recognition}, 74:406--421, 2018.

\bibitem{ergen2019unsupervisedlstm}
Tolga Ergen and Suleyman~Serdar Kozat.
\newblock Unsupervised anomaly detection with {LSTM} neural networks.
\newblock {\em IEEE transactions on neural networks and learning systems},
  31(8):3127--3141, 2019.

\bibitem{ester1996dbscan}
Martin Ester, Hans-Peter Kriegel, J{\"o}rg Sander, Xiaowei Xu, et~al.
\newblock A density-based algorithm for discovering clusters in large spatial
  databases with noise.
\newblock In {\em Kdd}, volume~96, pages 226--231, 1996.

\bibitem{farnaaz2016random}
Nabila Farnaaz and MA~Jabbar.
\newblock Random forest modeling for network intrusion detection system.
\newblock {\em Procedia Computer Science}, 89(1):213--217, 2016.

\bibitem{feng2014mining}
Wenying Feng, Qinglei Zhang, Gongzhu Hu, and Jimmy~Xiangji Huang.
\newblock Mining network data for intrusion detection through combining svms
  with ant colony networks.
\newblock {\em Future Generation Computer Systems}, 37:127--140, 2014.

\bibitem{feng2006csoacn}
Yong Feng, Jiang Zhong, Chun-xiao Ye, and Zhong-fu Wu.
\newblock Clustering based on self-organizing ant colony networks with
  application to intrusion detection.
\newblock In {\em Sixth International Conference on Intelligent Systems Design
  and Applications}, volume~2, pages 1077--1080. IEEE, 2006.

\bibitem{franti2019howmuchkmeans}
Pasi Fr{\"a}nti and Sami Sieranoja.
\newblock How much can k-means be improved by using better initialization and
  repeats?
\newblock {\em Pattern Recognition}, 93:95--112, 2019.

\bibitem{gama2014surveyconceptdrift}
Jo{\~a}o Gama, Indr{\.e} {\v{Z}}liobait{\.e}, Albert Bifet, Mykola Pechenizkiy,
  and Abdelhamid Bouchachia.
\newblock A survey on concept drift adaptation.
\newblock {\em ACM computing surveys (CSUR)}, 46(4):1--37, 2014.

\bibitem{gormley2015Elasticsearch}
Clinton Gormley and Zachary Tong.
\newblock {\em Elasticsearch: the definitive guide: a distributed real-time
  search and analytics engine}.
\newblock "O'Reilly Media, Inc.", 2015.

\bibitem{gupta2015kibana}
Yuvraj Gupta.
\newblock {\em Kibana essentials}.
\newblock Packt Publishing Ltd, 2015.

\bibitem{habeeb2019realstreamingsurvey}
Riyaz Ahamed~Ariyaluran Habeeb, Fariza Nasaruddin, Abdullah Gani, Ibrahim
  Abaker~Targio Hashem, Ejaz Ahmed, and Muhammad Imran.
\newblock Real-time big data processing for anomaly detection: A survey.
\newblock {\em International Journal of Information Management}, 45:289--307,
  2019.

\bibitem{hamalainen2021improvingscalablekmeans}
Joonas H{\"a}m{\"a}l{\"a}inen, Tommi K{\"a}rkk{\"a}inen, and Tuomo Rossi.
\newblock Improving scalable k-means++.
\newblock {\em Algorithms}, 14(1):6, 2021.

\bibitem{hariri2018extendedIF}
Sahand Hariri, Matias~Carrasco Kind, and Robert~J Brunner.
\newblock Extended isolation forest.
\newblock {\em arXiv preprint arXiv:1811.02141}, 2018.

\bibitem{JH10}
Ben He, Jimmy~Xiangji Huang, and Xiaofeng Zhou.
\newblock Modeling term proximity for probabilistic information retrieval
  models.
\newblock {\em Information Sciences}, 181(14):3017--3031, 2011.

\bibitem{hossain2019distortion}
Md~Tahmid Hossain, Shyh~Wei Teng, Dengsheng Zhang, Suryani Lim, and Guojun Lu.
\newblock Distortion robust image classification using deep convolutional
  neural network with discrete cosine transform.
\newblock In {\em 2019 IEEE International Conference on Image Processing
  (ICIP)}, pages 659--663. IEEE, 2019.

\bibitem{JH1}
Xiangji Huang and Qinmin Hu.
\newblock A bayesian learning approach to promoting diversity in ranking for
  biomedical information retrieval.
\newblock In {\em Proceedings of the 32nd International {ACM} {SIGIR}
  Conference on Research and Development in Information Retrieval,}, pages
  307--314, 2009.

\bibitem{JH2}
Xiangji Huang, Fuchun Peng, Dale Schuurmans, Nick Cercone, and Stephen~E.
  Robertson.
\newblock Applying machine learning to text segmentation for information
  retrieval.
\newblock {\em Information Retrieval}, 6(3-4):333--362, 2003.

\bibitem{JH4}
Xiangji Huang, Ming Zhong, and Luo Si.
\newblock {York University} at {TREC} 2005: Genomics track.
\newblock In {\em Proceedings of the Fourteenth Text REtrieval Conference,
  {TREC}}, 2005.

\bibitem{javaid2016deep2}
Ahmad Javaid, Quamar Niyaz, Weiqing Sun, and Mansoor Alam.
\newblock A deep learning approach for network intrusion detection system.
\newblock In {\em Proceedings of the 9th EAI International Conference on
  Bio-inspired Information and Communications Technologies (formerly
  BIONETICS)}, pages 21--26, 2016.

\bibitem{laskar2020query}
Md~Tahmid~Rahman Laskar, Enamul Hoque, and Jimmy Huang.
\newblock Query focused abstractive summarization via incorporating query
  relevance and transfer learning with transformer models.
\newblock In {\em Canadian Conference on Artificial Intelligence}, pages
  342--348. Springer, 2020.

\bibitem{laskar2020utilizing}
Md~Tahmid~Rahman Laskar, Enamul Hoque, and Jimmy~Xiangji Huang.
\newblock Utilizing bidirectional encoder representations from transformers for
  answer selection.
\newblock {\em arXiv preprint arXiv:2011.07208}, 2020.

\bibitem{laskar2020wslds}
Md~Tahmid~Rahman Laskar, Enamul Hoque, and Xiangji Huang.
\newblock {WSL-DS}: Weakly supervised learning with distant supervision for
  query focused multi-document abstractive summarization.
\newblock In {\em Proceedings of the 28th International Conference on
  Computational Linguistics}, pages 5647--5654, 2020.

\bibitem{laskar-LREC}
Md~Tahmid~Rahman Laskar, Jimmy~Xiangji Huang, and Enamul Hoque.
\newblock Contextualized embeddings based transformer encoder for sentence
  similarity modeling in answer selection task.
\newblock In {\em Proceedings of The 12th Language Resources and Evaluation
  Conference}, pages 5505--5514, 2020.

\bibitem{leung2005unsupervisedclustering}
Kingsly Leung and Christopher Leckie.
\newblock Unsupervised anomaly detection in network intrusion detection using
  clusters.
\newblock In {\em Proceedings of the Twenty-eighth Australasian conference on
  Computer Science-Volume 38}, pages 333--342, 2005.

\bibitem{liao2002use}
Yihua Liao and V~Rao Vemuri.
\newblock Use of k-nearest neighbor classifier for intrusion detection.
\newblock {\em Computers \& security}, 21(5):439--448, 2002.

\bibitem{liu2008isolation}
Fei~Tony Liu, Kai~Ming Ting, and Zhi-Hua Zhou.
\newblock Isolation forest.
\newblock In {\em 2008 Eighth IEEE International Conference on Data Mining},
  pages 413--422. IEEE, 2008.

\bibitem{liu2012isolation}
Fei~Tony Liu, Kai~Ming Ting, and Zhi-Hua Zhou.
\newblock Isolation-based anomaly detection.
\newblock {\em ACM Transactions on Knowledge Discovery from Data (TKDD)},
  6(1):1--39, 2012.

\bibitem{JH7}
Yang Liu, Aijun An, and Xiangji Huang.
\newblock Boosting prediction accuracy on imbalanced datasets with {SVM}
  ensembles.
\newblock In {\em Proceedings of the 10th Pacific-Asia Conference on Knowledge
  Discovery and Data Mining, {PAKDD}}, pages 107--118, 2006.

\bibitem{JH5}
Yang Liu, Xiangji Huang, Aijun An, and Xiaohui Yu.
\newblock {ARSA:} a sentiment-aware model for predicting sales performance
  using blogs.
\newblock In {\em Proceedings of the 30th International {ACM} {SIGIR}
  Conference on Research and Development in Information Retrieval}, pages
  607--614, 2007.

\bibitem{tcps4}
Meiyi Ma, Sarah~M Preum, Mohsin~Y Ahmed, William T{\"a}rneberg, Abdeltawab
  Hendawi, and John~A Stankovic.
\newblock Data sets, modeling, and decision making in smart cities: A survey.
\newblock {\em ACM Transactions on Cyber-Physical Systems}, 4(2):1--28, 2019.

\bibitem{meng2016mllib}
Xiangrui Meng, Joseph Bradley, Burak Yavuz, Evan Sparks, Shivaram Venkataraman,
  Davies Liu, Jeremy Freeman, DB~Tsai, Manish Amde, Sean Owen, et~al.
\newblock Mllib: Machine learning in apache spark.
\newblock {\em The Journal of Machine Learning Research}, 17(1):1235--1241,
  2016.

\bibitem{mukherjee2012intrusion}
Saurabh Mukherjee and Neelam Sharma.
\newblock Intrusion detection using naive bayes classifier with feature
  reduction.
\newblock {\em Procedia Technology}, 4:119--128, 2012.

\bibitem{mukkamala2002intrusion}
Srinivas Mukkamala, Guadalupe Janoski, and Andrew Sung.
\newblock Intrusion detection using neural networks and support vector
  machines.
\newblock In {\em Proceedings of the 2002 International Joint Conference on
  Neural Networks. IJCNN'02 (Cat. No. 02CH37290)}, volume~2, pages 1702--1707.
  IEEE, 2002.

\bibitem{munir2019localized}
Ahnaf Munir, Md~Tahmid~Rahman Laskar, Md~Sakhawat Hossen, and Salimur
  Choudhury.
\newblock A localized fault tolerant load balancing algorithm for rfid systems.
\newblock {\em Journal of Ambient Intelligence and Humanized Computing},
  10(11):4305--4317, 2019.

\bibitem{tcps5}
Sirajum Munir, Hao-Tsung Yang, Shan Lin, SM~Shahriar Nirjon, Chen Lin, Enamul
  Hoque, John~A Stankovic, and Kamin Whitehouse.
\newblock Reliable communication and latency bound generation in wireless
  cyber-physical systems.
\newblock {\em ACM Transactions on Cyber-Physical Systems}, 4(2):1--26, 2019.

\bibitem{myerson2001areaAUC1}
Joel Myerson, Leonard Green, and Missaka Warusawitharana.
\newblock Area under the curve as a measure of discounting.
\newblock {\em Journal of the experimental analysis of behavior},
  76(2):235--243, 2001.

\bibitem{newaz2019healthguard}
AKM~Iqtidar Newaz, Amit~Kumar Sikder, Mohammad~Ashiqur Rahman, and A~Selcuk
  Uluagac.
\newblock Healthguard: A machine learning-based security framework for smart
  healthcare systems.
\newblock In {\em 2019 Sixth International Conference on Social Networks
  Analysis, Management and Security (SNAMS)}, pages 389--396. IEEE, 2019.

\bibitem{otoum2018adaptively}
Safa Otoum, Burak Kantarci, and Hussein Mouftah.
\newblock Adaptively supervised and intrusion-aware data aggregation for
  wireless sensor clusters in critical infrastructures.
\newblock In {\em 2018 IEEE international conference on communications (ICC)},
  pages 1--6. IEEE, 2018.

\bibitem{otoum2019empowering}
Safa Otoum, Burak Kantarci, and Hussein Mouftah.
\newblock Empowering reinforcement learning on big sensed data for intrusion
  detection.
\newblock In {\em Icc 2019-2019 IEEE international conference on communications
  (ICC)}, pages 1--7. IEEE, 2019.

\bibitem{otoum2020comparative}
Safa Otoum, Burak Kantarci, and Hussein Mouftah.
\newblock A comparative study of ai-based intrusion detection techniques in
  critical infrastructures.
\newblock {\em arXiv preprint arXiv:2008.00088}, 2020.

\bibitem{otoum2017detectionrandom}
Safa Otoum, Burak Kantarci, and Hussein~T Mouftah.
\newblock Detection of known and unknown intrusive sensor behavior in critical
  applications.
\newblock {\em IEEE Sensors Letters}, 1(5):1--4, 2017.

\bibitem{otoum2020novelensemble}
Safa Otoum, Burak Kantarci, and Hussein~T Mouftah.
\newblock A novel ensemble method for advanced intrusion detection in wireless
  sensor networks.
\newblock In {\em Icc 2020-2020 ieee international conference on communications
  (icc)}, pages 1--6. IEEE, 2020.

\bibitem{preiss2008datastructure}
Bruno~R Preiss.
\newblock {\em Data structures and algorithms with object-oriented design
  patterns in C++}.
\newblock John Wiley \& Sons, 2008.

\bibitem{pruessner2003twoAUC2}
Jens~C Pruessner, Clemens Kirschbaum, Gunther Meinlschmid, and Dirk~H
  Hellhammer.
\newblock Two formulas for computation of the area under the curve represent
  measures of total hormone concentration versus time-dependent change.
\newblock {\em Psychoneuroendocrinology}, 28(7):916--931, 2003.

\bibitem{servin2005multireinforcement}
Arturo Servin and Daniel Kudenko.
\newblock Multi-agent reinforcement learning for intrusion detection.
\newblock In {\em Adaptive Agents and Multi-Agent Systems III. Adaptation and
  Multi-Agent Learning}, pages 211--223. Springer, 2005.

\bibitem{shi2006unsupervisedmulcross}
Tao Shi and Steve Horvath.
\newblock Unsupervised learning with random forest predictors.
\newblock {\em Journal of Computational and Graphical Statistics},
  15(1):118--138, 2006.

\bibitem{shone2018deep1}
Nathan Shone, Tran~Nguyen Ngoc, Vu~Dinh Phai, and Qi~Shi.
\newblock A deep learning approach to network intrusion detection.
\newblock {\em IEEE Transactions on Emerging Topics in Computational
  Intelligence}, 2(1):41--50, 2018.

\bibitem{tcps2}
Yasser Shoukry, Michelle Chong, Masashi Wakaiki, Pierluigi Nuzzo, Alberto
  Sangiovanni-Vincentelli, Sanjit~A Seshia, Joao~P Hespanha, and Paulo Tabuada.
\newblock Smt-based observer design for cyber-physical systems under sensor
  attacks.
\newblock {\em ACM Transactions on Cyber-Physical Systems}, 2(1):1--27, 2018.

\bibitem{shukla2015Elasticsearch}
Vishal Shukla.
\newblock {\em Elasticsearch for Hadoop}.
\newblock Packt Publishing Ltd, 2015.

\bibitem{shvachko2010hdfs}
Konstantin Shvachko, Hairong Kuang, Sanjay Radia, and Robert Chansler.
\newblock The hadoop distributed file system.
\newblock In {\em 2010 IEEE 26th symposium on mass storage systems and
  technologies (MSST)}, pages 1--10. IEEE, 2010.

\bibitem{sommer2010outsideML}
Robin Sommer and Vern Paxson.
\newblock Outside the closed world: On using machine learning for network
  intrusion detection.
\newblock In {\em 2010 IEEE symposium on security and privacy}, pages 305--316.
  IEEE, 2010.

\bibitem{sotiris2010anomalySVMbayesian}
Vasilis~A Sotiris, W~Tse Peter, and Michael~G Pecht.
\newblock Anomaly detection through a bayesian support vector machine.
\newblock {\em IEEE Transactions on Reliability}, 59(2):277--286, 2010.

\bibitem{terzi2017bigdata}
Duygu~Sinanc Terzi, Ramazan Terzi, and Seref Sagiroglu.
\newblock Big data analytics for network anomaly detection from netflow data.
\newblock In {\em 2017 International Conference on Computer Science and
  Engineering (UBMK)}, pages 592--597. IEEE, 2017.

\bibitem{thottan2003anomalyIP}
Marina Thottan and Chuanyi Ji.
\newblock Anomaly detection in ip networks.
\newblock {\em IEEE Transactions on signal processing}, 51(8):2191--2204, 2003.

\bibitem{tian2014anomalysupervisedlabel}
Jing Tian, Michael~H Azarian, and Michael Pecht.
\newblock Anomaly detection using self-organizing maps-based k-nearest neighbor
  algorithm.
\newblock In {\em Proceedings of the European Conference of the Prognostics and
  Health Management Society}, pages 1--9. Citeseer, 2014.

\bibitem{tiyas2014reinforced}
Indah Yulia~Prafitaning Tiyas, Ali~Ridho Barakbah, Tri Harsono, and Amang
  Sudarsono.
\newblock Reinforced intrusion detection using pursuit reinforcement
  competitive learning.
\newblock {\em EMITTER International Journal of Engineering Technology},
  2(1):39--49, 2014.

\bibitem{lof}
Luan Tran, Liyue Fan, and Cyrus Shahabi.
\newblock Distance-based outlier detection in data streams.
\newblock {\em Proceedings of the VLDB Endowment}, 9(12):1089--1100, 2016.

\bibitem{vaswani2017attention}
Ashish Vaswani, Noam Shazeer, Niki Parmar, Jakob Uszkoreit, Llion Jones,
  Aidan~N Gomez, {\L}ukasz Kaiser, and Illia Polosukhin.
\newblock Attention is all you need.
\newblock In {\em Proceedings of the 31st International Conference on Neural
  Information Processing Systems}, pages 6000--6010, 2017.

\bibitem{linkedin}
James Verbus and contributors.
\newblock linkedin-isolation-forest.
\newblock \url{https://github.com/linkedin/isolation-forest}, 2019.

\bibitem{vinayakumar2017cnn}
R~Vinayakumar, KP~Soman, and Prabaharan Poornachandran.
\newblock Applying convolutional neural network for network intrusion
  detection.
\newblock In {\em 2017 International Conference on Advances in Computing,
  Communications and Informatics (ICACCI)}, pages 1222--1228. IEEE, 2017.

\bibitem{tcps3}
Chang Wang, Yongxin Zhu, Weiwei Shi, Victor Chang, Pandi Vijayakumar, Bin Liu,
  Yishu Mao, Jiabao Wang, and Yiping Fan.
\newblock A dependable time series analytic framework for cyber-physical
  systems of iot-based smart grid.
\newblock {\em ACM Transactions on Cyber-Physical Systems}, 3(1):1--18, 2018.

\bibitem{white2012hadoop}
Tom White.
\newblock {\em Hadoop: The definitive guide}.
\newblock "O'Reilly Media, Inc.", 2012.

\bibitem{wold1987principal}
Svante Wold, Kim Esbensen, and Paul Geladi.
\newblock Principal component analysis.
\newblock {\em Chemometrics and intelligent laboratory systems}, 2(1-3):37--52,
  1987.

\bibitem{wu2020anomaly}
Di~Wu, Hanlin Zhu, Yongxin Zhu, Victor Chang, Cong He, Ching-Hsien Hsu, Hui
  Wang, Songlin Feng, Li~Tian, and Zunkai Huang.
\newblock Anomaly detection based on rbm-lstm neural network for cps in
  advanced driver assistance system.
\newblock {\em ACM Transactions on Cyber-Physical Systems}, 4(3):1--17, 2020.

\bibitem{xu2005reinforcement}
Xin Xu and Tao Xie.
\newblock A reinforcement learning approach for host-based intrusion detection
  using sequences of system calls.
\newblock In {\em International Conference on Intelligent Computing}, pages
  995--1003. Springer, 2005.

\bibitem{yamanishi2004linehttp}
Kenji Yamanishi, Jun-Ichi Takeuchi, Graham Williams, and Peter Milne.
\newblock On-line unsupervised outlier detection using finite mixtures with
  discounting learning algorithms.
\newblock {\em Data Mining and Knowledge Discovery}, 8(3):275--300, 2004.

\bibitem{titicacasparkiforest}
Fangzhou Yang and contributors.
\newblock spark-iforest.
\newblock \url{https://github.com/titicaca/spark-iforest}, 2018.

\bibitem{yin2017deep3}
Chuanlong Yin, Yuefei Zhu, Jinlong Fei, and Xinzheng He.
\newblock A deep learning approach for intrusion detection using recurrent
  neural networks.
\newblock {\em IEEE Access}, 5:21954--21961, 2017.

\bibitem{JH3}
Xiaoshi Yin, Jimmy~Xiangji Huang, Zhoujun Li, and Xiaofeng Zhou.
\newblock A survival modeling approach to biomedical search result
  diversification using wikipedia.
\newblock {\em {IEEE} Transactions on Knowledge and Data Engineering},
  25(6):1201--1212, 2013.

\bibitem{tcps7}
Sze~Zheng Yong, Minghui Zhu, and Emilio Frazzoli.
\newblock Switching and data injection attacks on stochastic cyber-physical
  systems: Modeling, resilient estimation, and attack mitigation.
\newblock {\em ACM Transactions on Cyber-Physical Systems}, 2(2):1--2, 2018.

\bibitem{JH6}
Xiaohui Yu, Yang Liu, Xiangji Huang, and Aijun An.
\newblock Mining online reviews for predicting sales performance: {A} case
  study in the movie domain.
\newblock {\em {IEEE} Transactions on Knowledge and Data Engineering},
  24(4):720--734, 2012.

\bibitem{zaharia2016apachespark}
Matei Zaharia, Reynold~S Xin, Patrick Wendell, Tathagata Das, Michael Armbrust,
  Ankur Dave, Xiangrui Meng, Josh Rosen, Shivaram Venkataraman, Michael~J
  Franklin, et~al.
\newblock Apache spark: a unified engine for big data processing.
\newblock {\em Communications of the ACM}, 59(11):56--65, 2016.

\bibitem{zobaed2019clustcrypt}
SM~Zobaed, Sahan Ahmad, Raju Gottumukkala, and Mohsen~Amini Salehi.
\newblock Clustcrypt: Privacy-preserving clustering of unstructured big data in
  the cloud.
\newblock In {\em 2019 IEEE 21st International Conference on High Performance
  Computing and Communications; IEEE 17th International Conference on Smart
  City; IEEE 5th International Conference on Data Science and Systems
  (HPCC/SmartCity/DSS)}, pages 609--616. IEEE, 2019.

\end{thebibliography}

\end{document}